\documentclass[iop,apj]{emulateapj}   

\usepackage{ulem}
\usepackage{amsmath}

\usepackage{epsfig}
\usepackage[usenames, dvipsnames]{color}
\usepackage{natbib}
\usepackage{enumerate}
\usepackage{graphicx}
\usepackage{apjfonts}
\usepackage{multirow}
\usepackage{hyperref}
\usepackage[all]{hypcap}    
\hypersetup{
    colorlinks,
    citecolor=blue,
    filecolor=blue,
    linkcolor=blue,
    urlcolor=blue
}

\newcommand{\fesc}{\ifmmode{f_{\rm esc}}\else{$f_{\rm esc}$}\fi}
\newcommand{\fescs}{\ifmmode{f_{\rm esc}^\star}\else{$f_{\rm esc}^\star$}\fi}
\newcommand{\kms}{\ifmmode{{\;\rm km~s^{-1}}}\else{km~s$^{-1}$}\fi}
\newcommand{\fgas}{\ifmmode{{f_{\rm gas}}}\else{$f_{\rm gas}$}\fi}
\newcommand{\cubecm}{\ifmmode{{\rm cm^{-3}}}\else{cm$^{-3}$}\fi}
\newcommand{\ztwo}{\ifmmode{{\rm [Z_2/H]}}\else{[Z$_2$/H]}\fi}
\newcommand{\zthree}{\ifmmode{{\rm [Z_3/H]}}\else{[Z$_3$/H]}\fi}
\newcommand{\lsim}{\lower0.3em\hbox{$\,\buildrel <\over\sim\,$}}
\newcommand{\gsim}{\lower0.3em\hbox{$\,\buildrel >\over\sim\,$}}

\newcommand{\eavg}{\ifmmode{\langle E_\gamma \rangle}\else{$\langle E_\gamma \rangle$}\fi}

\newcommand{\enzo}{{\sl Enzo}}
\newcommand{\moray}{{\sl Enzo+Moray}}
\newcommand{\Ms}{\ifmmode{\textrm{M}_\odot}\else{$M_\odot$}\fi}
\newcommand{\vrms}{\ifmmode{v_{\rm rms}}\else{$v_{\rm rms}$}\fi}

\newcommand{\hh}{H$_2$}

\newcommand{\tvir}{\ifmmode{T_{\rm{vir}}}\else{$T_{\rm{vir}}$}\fi}
\newcommand{\mvir}{\ifmmode{M_{\rm{vir}}}\else{$M_{\rm{vir}}$}\fi}
\newcommand{\rvir}{\ifmmode{r_{\rm{vir}}}\else{$r_{\rm{vir}}$}\fi}

\newcommand{\jj}{\ifmmode{J_{21}}\else{$J_{21}$}\fi}
\newcommand{\flw}{\ifmmode{F_{LW}}\else{$F_{LW}$}\fi}
\newcommand{\kph}{\ifmmode{k_{\rm ph}}\else{$k_{\rm ph}$}\fi}

\newcommand{\zsun}{\ifmmode{\rm\,Z_\odot}\else{$\rm\,Z_\odot$}\fi}

\newcommand{\hi}{H {\sc i}}
\newcommand{\hii}{H {\sc ii}}

\newcommand\unit[1]{\; \textrm{#1}}

\begin{document}

\shorttitle{FIRST GALAXY PROPERTIES AND UV ESCAPE FRACTIONS}
\shortauthors{XU ET AL.}

\title{Galaxy Properties and UV Escape Fractions during Epoch of
  Reionization: Results from the {\it Renaissance Simulations}}


\author{Hao Xu\altaffilmark{1},
John H. Wise\altaffilmark{2},    
Michael L. Norman\altaffilmark{1},
Kyungjin Ahn\altaffilmark{3}
and Brian W. O'Shea\altaffilmark{4}}
\altaffiltext{1}{Center for Astrophysics and Space Sciences,
  University of California, San Diego, 9500 Gilman Drive, La Jolla, CA
  92093; hxu@ucsd.edu, mlnorman@ucsd.edu}
\altaffiltext{2}{Center for Relativistic Astrophysics, School of
  Physics, Georgia Institute of Technology, 837 State Street, Atlanta,
  GA 30332; jwise@gatech.edu}
\altaffiltext{3}{Department of Earth Science Education, Chosun University,
Gwangju 501-759, Korea; kjahn@chosun.ac.kr}
\altaffiltext{4}{Department of Computational Mathematics, Science and
  Engineering, Department of Physics and Astronomy, and National
  Superconducting Cyclotron Laboratory, Michigan State University,
East Lansing, MI 48824; oshea@msu.edu}

\begin{abstract}

  Cosmic reionization is thought to be primarily fueled by the first
  generations of galaxies.  We examine their stellar and gaseous
  properties, focusing on the star formation rates and the escape of
  ionizing photons, as a function of halo mass, redshift, and
  environment using the full suite of the {\it Renaissance
    Simulations} with an eye to provide better inputs to global
  reionization simulations.  This suite, carried out with the adaptive
  mesh refinement code Enzo, is unprecedented in terms of their size
  and physical ingredients.  The simulations probe overdense, average,
  and underdense regions of the universe of several hundred comoving
  Mpc$^3$, each yielding a sample of over 3,000 halos in the mass
  range $10^7 - 10^{9.5}~\Ms$ at their final redshifts of 15, 12.5,
  and 8, respectively.  In the process, we simulate the effects of
  radiative and supernova feedback from 5,000 to 10,000 metal-free
  (Population III) stars in each simulation.  We find that halos as
  small as $10^7~\Ms$ are able to form stars due to metal-line cooling
  from earlier enrichment by massive Population III stars.  However,
  we find such halos do not form stars continuously.  Using our large
  sample, we find that the galaxy-halo occupation fraction drops from
  unity at virial masses above $10^{8.5}~\Ms$ to $\sim$50\% at $10^8
  ~\Ms$ and $\sim$10\% at $10^7~\Ms$, quite independent of redshift and
  region.  Their average ionizing escape fraction is $\sim$5\% in the
  mass range $10^8 - 10^9~\Ms$ and increases with decreasing halo mass
  below this range, reaching 40--60\% at $10^7~\Ms$.  Interestingly,
  we find that the escape fraction varies between 10--20\% in halos
  with virial masses $\sim 3 \times 10^9~\Ms$.  Taken together, our
  results confirm the importance of the smallest galaxies as sources
  of ionizing radiation contributing to the reionization of the
  universe.

\end{abstract}


\section{Introduction}
\label{sec:intro}
It is believed that low-mass galaxies at $z \ga 6$ are the primary
sources of hydrogen ionizing photons to complete reionization.  These
galaxies naturally have low signal-to-noise ratios with current
telescopes because they are distant and intrinsically dim.
Nevertheless, recent observational campaigns have provided valuable
constraints on the nature of the first galaxies and their role during
reionization.  The Hubble Ultra Deep Field (HUDF) 2009 and 2012
campaigns \citep{Ellis13} can probe galaxies with rest frame UV
magnitude as low as $\sim -18$ at $z \ga 7$ and as distant as $z \simeq
11$ \citep{McLure11, Zheng12, Coe13, Finkelstein15, Oesch16}.

It is clear that UV ionizing photons from these observed ``bright
galaxies'' are not enough to fully ionize the universe by $z=6$
\citep{Robertson13}, as implied by quasar observations \citep{Fan06},
thus fainter galaxies and other ionizing sources (e.g. accreting black
holes) are needed to provide the remaining ionizing
photons. \citet{Robertson15}, by extrapolating the UV luminosity
function (LF) with a steep faint end slope $< -2$
\citep[e.g.][]{Bouwens11, Bouwens15, McLure13}, have shown that the LF
must extend to $M_{\rm UV} \sim -13$ to be consistent with the
integrated Thomson optical depth $\tau_{\rm es} = 0.058 \pm 0.012$
measured by the {\it Planck} satellite \citep{Planck16_Reion}.  The
latest analysis of the \textit{Frontier Fields} have suggested that
the LF shows no sign of flattening down to $M_{\rm UV} \simeq -13$
\citep{Atek15, Livermore16}.  Galaxies with such magnitudes have
stellar masses as small as 10$^{6}$ M$_\odot$ in halos with masses $M
\sim 10^{8}~\Ms$, providing sufficient UV radiation to complete and
maintain reionization.  The uncertainty in $\tau_{\rm es}$ allows for
some contribution from star formation occurring in minihalos with
masses $M \la 10^8~\Ms$ \citep{Ahn12, Salvadori14}.  Furthermore,
faint active galactic nuclei have been detected in ``normal'' star
forming galaxies with $M_{\rm UV}$ reaching up to --18.5
\citep{Giallongo15}, and they may contribute a non-negligible fraction
to the ionizing photon budget \citep{Madau15}.

This unseen population of even fainter and, perhaps, more abundant
galaxies will eventually be detected by next-generation telescopes
such the \textit{James Webb Space Telescope} \citep[JWST, launch date
2018;][]{Gardner06} and 30-meter class ground-based 
telescopes\footnote{European Extremely Large Telescope \citep[E-ELT, 39-m,
  completion date 2024;][]{ELT}, Giant Magellan Telescope \citep[GMT,
  24.5-m, completion date 2020;][]{GMT}, Thirty Meter Telescope
  \citep[TMT, 30-m, completion date 2022;][]{TMT}}.  However, the
growth of these small galaxies can be complicated by feedback from
both metal-free (Population III; Pop III) and metal-enriched stars,
where \hii~regions and supernovae (SNe) can drive outflows larger than
the escape velocity of their host halo \citep{Whalen04, Whalen08_SN,
  Kitayama04, Kitayama05, Abel07}, leaving behind a gas-poor halo that
only recovers by cosmological accretion after tens of Myr
\citep{Wise08_Gal, Greif10, Wise12a, Muratov13b, Jeon14}.  On the
other hand, these SNe also pre-enrich the gas that ultimately
assembles the first galaxies to $10^{-4} - 10^{-3} \zsun$
\citep{Bromm03_SN, Wise08_Gal, Karlsson08, Greif10, Wise12a}.  Prior
to cosmological reionization, galaxies can then form in DM halos as
small as $10^7~\Ms$.  Low-mass ($V_c = \sqrt{GM_{\rm vir}/R_{\rm vir}}
\la 30 \kms$) galaxies may provide $\sim$40\% of the ionizing photons
to reionization, eventually becoming photo-suppressed as reionization
ensues \citep{Wise14}.  A small fraction (5--15\%) of these first
galaxies may survive until the present day \citep{Gnedin06}, and
ultra-faint dwarf galaxies (UFDs) discovered in the Sloan Digital Sky
Survey (SDSS) that surround the Milky Way could be the fossils of this
subset of the first galaxies, providing a way to estimate the
abundance of dwarf galaxies during the epoch of reionization
\citep{Bullock00, Salvadori09, Bovill11b, Weisz14, Boylan14, Wheeler15}.


Provided that there is sufficient star formation during reionization,
the next question is the fraction of ionizing photons, $f_{\rm esc}$,
that can escape their host halos into the intergalactic medium (IGM).
This quantity is difficult to measure both observationally and
theoretically.  It is nearly impossible to detect Lyman continuum
(LyC) emission at $z > 4$ because Lyman limit systems become much more
abundant with increasing redshift \citep{Inoue08}.  Detecting LyC
emission only becomes feasible at $z \sim 3$ when the IGM optical
depth is around unity.  Deep narrow-band galaxy spectroscopy and
imaging have detected LyC emission in numerous $z \sim 3$ galaxies
with $f_{\rm esc}$ values ranging from an upper limit of 7--9\% for
bright galaxies \citep{Siana15} to $10-30$\% for fainter
Lyman-$\alpha$ emitters \citep{Nestor13}, $33 \pm 7$\% for
``Lyman-continuum galaxies'' \citep{Cooke14}, and $\ge 50$\% for a
compact star-forming galaxy {\it Ion2} with a stellar mass
$\sim$$10^{9}~\Ms$ \citep{Vanzella16_fesc}.  However, these
observations are susceptible to foreground contamination from lower
redshift galaxies in the same line of sight \citep[e.g.][]{Vanzella12,
  Siana15, Grazian16}.

Simulations have suggested that the escape fraction is on the order of
a few percent for halos with masses $\ge$ 10$^{11}$ M$_\odot$
\citep{Dove00, Razoumov07, Gnedin08, Yajima11}.  Conversely, recent
post-processing results from the EAGLE simulation \citep{EAGLE}
estimated that \fesc{} is on the order 10--20\% in galaxies with star
formation rates $\mathrm{SFR} \ge 1 \Ms \unit{yr}^{-1}$ and increases
with SFR \citep{Sharma16}.  By redshift 6, they found that galaxies
with $M_{\rm UV} < -18$ may provide half of the photon budget to the
global ionizing emissivity.  If lower-mass galaxies have similar
escape fractions before reionization, there are not enough ionizing
photons that can escape galaxies to reionize the universe by $z=6$
\citep{Gnedin08}.  

The escape fraction from small galaxies is also under
debate. \citet{Wood00} argued that f$_{\rm esc}$ $\le$ 0.01, due to
the higher mean densities at high redshifts. Using idealized isolated
galaxy calculations, \citet{Fujita03} found that $\fesc{} \le 0.01$
from dwarf starburst disc galaxies with total masses between 10$^8$
and 10$^{10}$ M$_\odot$, and \citet{Paardekooper11} found similar
results for high-redshift disc galaxies with total masses of 10$^8$
and 10$^9$ M$_\odot$. In contrast, \citet{Ricotti00} found higher
escape fraction $\fesc{} \ge 0.1$ in high-redshift halos with masses
$M \le 10^7~\Ms$.  Such escape fraction values were then further
confirmed by several numerical simulations \citep{Wise09, Razoumov10,
  Yajima11, Paardekooper13, Paardekooper15, Ferrara13}.
\citet{Wise14}, using high resolution cosmological radiation
hydrodynamics simulations of early galaxies, showed that the mean
escape fraction of hydrogen ionizing photons decreases with increasing
halo mass. They found that the amount of ionizing photons per unit
mass escaping from a halo exhibits little evolution with a wide halo
mass range, from 10$^{6.75}$ to 10$^{8.75}$ M$_\odot$.  They concluded
that low-mass galaxies ($M_{\rm halo} \ge 10^8 M_{\odot}$) may produce
a significant amount of the ionizing photons escaping into the IGM at
$z \ge 10$ during cosmic reionization, suggesting that the faintest
galaxies (M$_{UV}$ $\ge$ --12) are very important in the early stage
of the epoch of reionization.

However, star formation in these low-mass galaxies is suppressed as
reionization progresses, and galaxies that are not susceptible to
photo-evaporation ($M_{\rm vir} \ga 10^{9} \Ms$) provide the remaining
ionizing radiation to complete reionization.  Post-process radiative
transfer calculations \citep{Ma15, Sharma16} and radiation
hydrodynamics simulations \citep{Kimm14, Gnedin16} that studied this
mass range found that \fesc{} is highly variable in a single galaxy
and has a large scatter for a given galaxy mass, halo mass, or UV
luminosity.  Time-averaged \fesc{} values between 3--15\% in halos
with $M_{\rm vir} \ga 10^{9} \Ms$, but any trends with halo mass or
redshift either weak or inconsistent between groups.  Further
complicating the issue, runaway and binary stars may boost \fesc{}
values by a few percent \citep{Conroy12, Kimm14, Ma16}.

A major caveat in the work of \citeauthor{Wise14} is that the
simulation volume of 1 comoving Mpc$^3$ is small and has no
large-scale variance in cosmological density distribution. There are
only 32 dwarf galaxies at the final redshift, and the simulation does
not capture galaxies forming in halos more massive than 10$^9$
M$_{\odot}$. So, in their paper they combined the data at different
redshifts to statistically evaluate the galaxy properties of star
formation and escape fraction by assuming that these properties are
independent of redshift and environment.

In this work, we characterize the abundance and escape fraction of
ionizing radiation from faint galaxies before cosmic reionization
utilizing a suite of zoom-in cosmological radiation hydrodynamics
simulations, named the {\it Renaissance Simulations}, that survey
three regions with varying large-scale overdensities.  These
quantities are extremely important in understanding the progression of
reionization.  Each simulation in this work improves the statistics of
the first galaxy properties in \citet{Wise14} by a factor of
$\sim$100.  We validate their scheme to combine data from different
times, confirm their results in different environments, and extend the
analysis to slightly more massive galaxies. In addition to elucidating
the ionizing photon budget during cosmic reionization, our work can
provide valuable constraints on the process of reionization.  We first
describe our simulation setup in Section \ref{sec:sims}. Then, in
Section \ref{sec:prop}, we present results on the overall baryonic
galaxy properties, on how reionization initially proceeds, and on the
similarity of galaxy properties in different large-scale environments.
We then describe the method used to calculate ionizing escape
fractions in each galaxy and present the variations of this fraction
in Section \ref{sec:fesc}.  Last, we discuss and conclude our findings
in Section \ref{sec:conclusions}.


\vspace{3em}
\section{Simulations}
\label{sec:sims}
We present results from the {\it Renaissance Simulations}, a
suite of zoom-in calculations that focus on high-redshift ($z \ge 8$)
galaxy formation and the ensuing reionization, which were originally presented in
\citet{OShea15}.  Each simulation encompasses a different large-scale
environment with a comoving volume of $\sim 200~\textrm{Mpc}^3$ and
includes metal-free and metal-enriched star formation and feedback.
These simulations self-consistently capture the formation of nearly
3,000 of the first
generations of galaxies, and we study their surrounding environment,
their baryonic properties, and the self-regulation of their star
formation.  More specifically, we focus on the photo-ionization and
photo-heating of the IGM and the role of the first galaxies during
reionization.  We have previously presented results from the most
overdense region (the ``Rare Peak'' simulation) in the {\it Renaissance Simulations}
on the Pop III stellar distribution \citep{Xu13}, X-rays from Pop III
binaries \citep{Xu14}, their 21-cm signal \citep{Ahn15}, their
scaling relations \citep{Chen14}, and the galaxy luminosity function \citep{OShea15}.

\subsection{Simulation setup}

\begin{figure}[t]
\centering
\includegraphics[width=1.0\columnwidth]{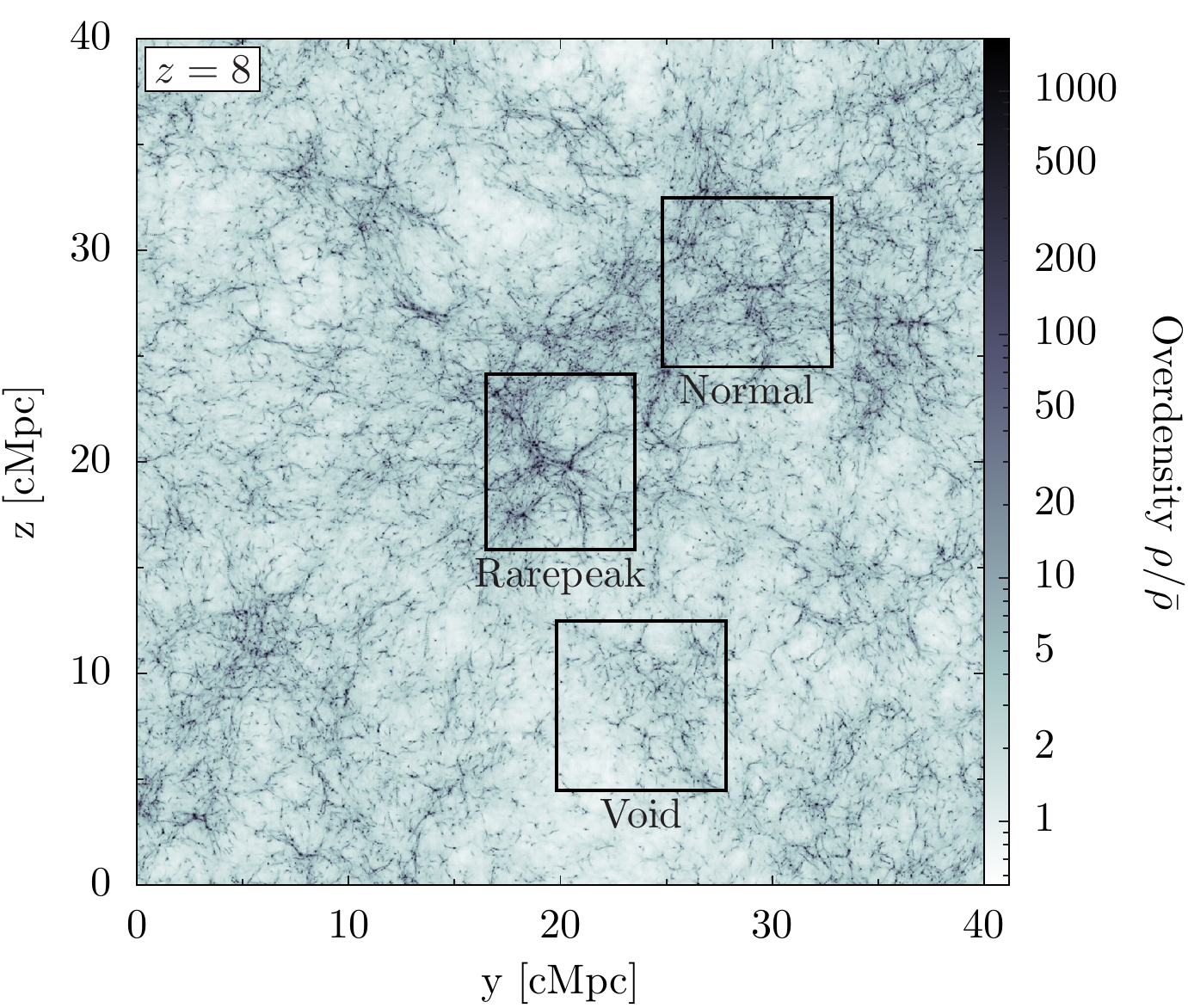}
\caption{Mass-weighted density projection of the (40 comoving Mpc)$^3$
  exploratory dark matter simulation at $z = 8$.  The survey volumes
  of the Rarepeak, Normal, and Void regions are outlined.  The
  Rarepeak region is centered on the most massive halo at $z=6$.  Due
  to projection effects, the normal region appears as dense as the
  Rarepeak, however its average overdensity is only 9\% higher than
  the mean matter density. \label{fig:fullbox}}
\end{figure}

We use the adaptive mesh refinement (AMR) cosmological hydrodynamics
code \enzo\ \citep{Bryan13}, along with its adaptive ray tracing module
\moray~\citep{Wise11} for the transport of ionizing
radiation, which is coupled to the hydrodynamics and chemistry in
Enzo.

All of the {\it Renaissance Simulations} are performed in the same
comoving volume of (40 Mpc)$^3$.  The initial conditions for this
volume are generated using \textsc{Music} \citep{Hahn11_MUSIC} with
second-order Lagrangian perturbations at $z=99$ using a 512$^3$ root
grid resolution.  We use the cosmological parameters from the 7-year
WMAP $\Lambda$CDM+SZ+LENS best fit \citep{Komatsu11}:
$\Omega_{M}=0.266$, $\Omega_{\Lambda} = 0.734$, $\Omega_{b}=0.0449$,
$h=0.71$, $\sigma_{8}=0.81$, and $n=0.963$.

It is computationally prohibitive to have the necessary parsec-scale
spatial resolution (and accompanying mass resolution), which is
required to marginally resolve star forming molecular clouds,
throughout the entire simulation volume.  We perform zoom-in
simulations on three selected regions, ranging from 220 to 430
comoving Mpc$^3$ with different overdensities, providing a mixture of
large-scale environments.  We first run a 512$^3$ N-body only
simulation to $z=6$. We then select an overdense region
(``Rarepeak''), a nearly mean density region (``Normal'') and an
underdense region (``Void''), which are displayed in Figure
\ref{fig:fullbox}. The selection of the survey volume and detailed
setup of the Rarepeak have been described in \citet{Xu13}, which is
centered on two $3 \times 10^{10}~\Ms$ halos at $z = 6$ with a survey
volume of $(3.8 \times 5.4 \times 6.6)\; \textrm{Mpc}^3$.  For both
the Normal and Void runs, we select comoving volumes of $(6.0 \times
6.0 \times 6.125)\; \textrm{Mpc}^3$ as the survey volumes.  We then
re-initialize all simulations, having the survey volume at the center,
with 3 more static nested grids to have an effective resolution of
4096$^3$ and an effective dark matter mass resolution of $2.9 \times
10^4$ $\Ms$ inside the highest static nested grid that encompasses the
survey volume.  During the course of the simulation, we allow a
maximum refinement level $l=12$, resulting in a maximal resolution of
19 comoving pc.  The refinement criteria employed are the same as in
\citet{Wise12a}, refining on baryon and dark matter overdensities of 4
and local Jeans length by at least 4 cells \citep{Truelove98} and is
restricted to the survey volumes.  While the Rarepeak simulation
adjusts the survey volume size during the simulation to contain only
the highest resolution dark matter particles of the highest static
nested grid, matter in the Normal and Void simulations is not fully
contained in a large-scale potential well and have significant
peculiar velocities, causing some of the high-resolution particles to
migrate out of the initial static grid.  Thus, we simplify the
simulation setup by restricting grid refinement to occur in the
initial high-resolution grid instead of its Lagrangian region.  We
stop the simulations of the Rarepeak, Normal, and Void regions at $z =
(15, 12.5, 8)$, respectively, because of the high computational cost
of the radiative transfer.  The Renaissance simulations were run on
the Blue Waters system at NCSA.  Each simulation used approximately
eight million core-hours.

\subsection{Lyman-Werner Background}

\begin{figure}[t]
\centering
\includegraphics[width=1.0\columnwidth]{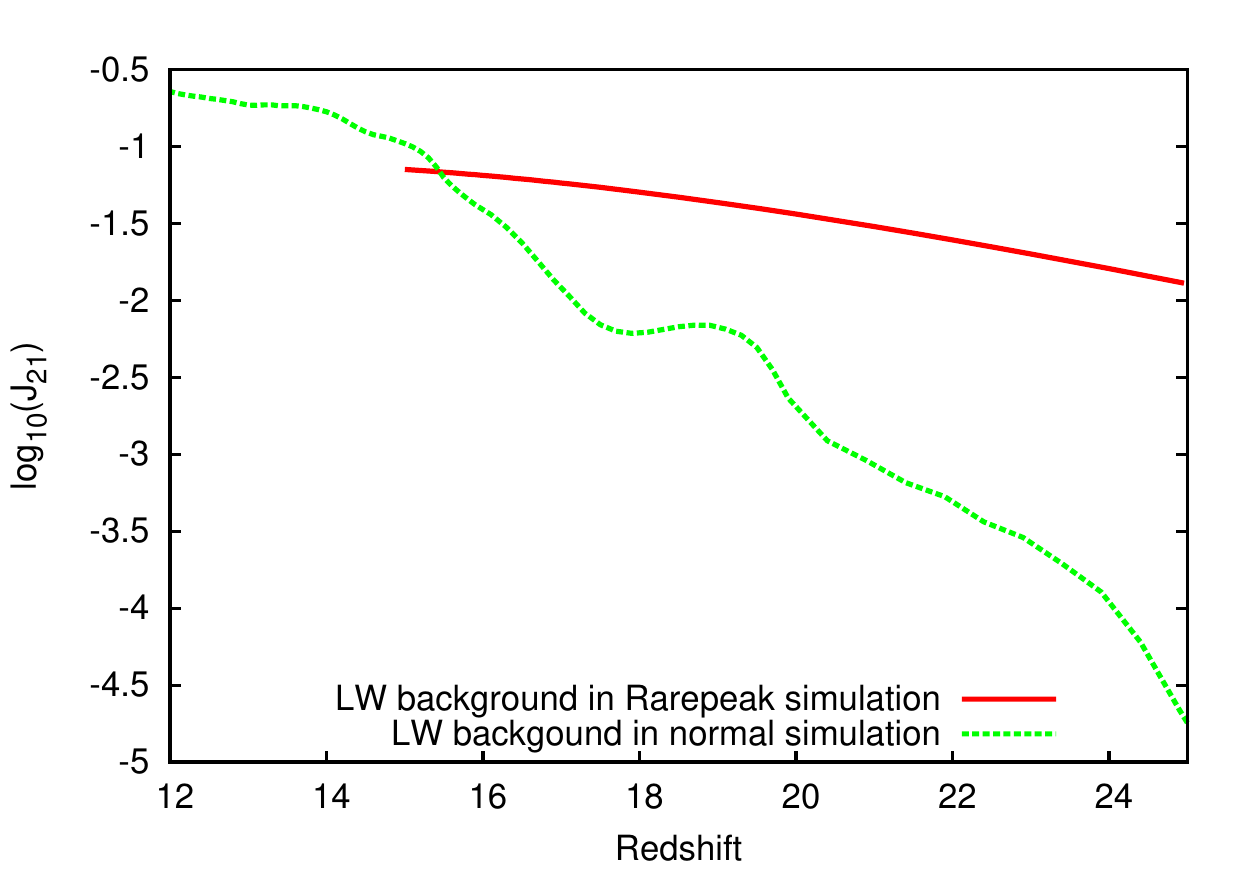}
\caption{Evolution of the intensity of the LW background used in the
  Normal (green) and Rarepeak (red) simulations.  The intensity in the
  Rarepeak is calculated with Equation (\ref{eqn:lwb}) and is
  higher because the simulation cannot capture the smallest
  star-forming halos \citep{Tegmark97, Machacek01} that have typical
  masses around $2 \times 10^5~\Ms$ in the absence of a LW
  background.  The Pop III stars that form in these small halos
  dominate the emissivity at very high redshifts and result the
  semi-analytic estimate (red) to be higher than the one based on
  the simulation (green).
  \label{fig:LW_background}}
\end{figure}

All of the simulations use the same chemistry, cooling, metal-free and
metal-enriched star formation, and radiative and supernova feedback
models, which are fully described in \citet{Wise12b} and \citet{Xu13}.
We model the H$_2$ dissociating radiation with an optically-thin,
inverse square profile, centered on all metal enriched and Pop III
star particles.  However, while we use the same model for Lyman-Werner
(LW) radiation from stellar sources, different LW backgrounds are used
in different simulations. LW radiation may significantly delay the Pop
III formation in low mass halos \citep{Machacek01, Wise07_UVB,
  OShea08}, and thus the enrichment history and the emergence of
metal-enriched stars in those halos.

The three simulations reported in the paper use three different LW
background models.  For the Rarepeak simulation, we use the
time-dependent LW optically thin radiation background used in
\citet{Wise12a}, which is based on the semi-analytical model of
\citet{Wise05}, updated with the 7-year WMAP parameters and optical
depth to Thomson scattering.  This model considers the LW
contributions of Pop III stars and galaxies, where the former
dominates the emissivity at $z \gsim 12$ before becoming suppressed
through \hh~photo-dissociation.  Because of the uncertainties in the
choice of the ionizing escape fraction and star formation efficiencies
in this model, we only apply it at higher redshifts ($z \gsim 12$)
before metal-enriched stars dominate the cosmic emissivity.  We use
the functional form of the background evolution in \citet{Wise12a},
\begin{equation}
\label{eqn:lwb}
\log_{10} J_{21}(z) = A + Bz + Cz^2 + Dz^3 + Ez^4, 
\end{equation}
where (A, B, C, D, E) = (-2.567, 0.4562, -0.02680, 5.882 $\times$
10$^{-4}$, -5.056 $\times$ 10$^{-6}$), and J$_{21}$ is the specific
intensity in units of 10$^{-21}$ erg s$^{-1}$ cm$^{-2}$ Hz$^{-1}$
sr$^{-1}$.  In the high density region of the Rarepeak simulation, the
LW radiation from local Pop III and metal-enriched star sources
dominate this LW background at redshifts as early as 20.

For the Normal region, we calculate the LW background self-consistently by
considering the actual evolution of sources inside the simulation
box. This choice is justified by the fact that the Normal region is
indeed a good representation of an average patch of the Universe. We
assume that sources outside the simulation box are uniformly
distributed, which is a good approximation because the spatial
fluctuations of the source distribution is smeared out when the
sources are located very far from the observing point. Then the
background $J_{21}(z)$ is given by the following calculation \citep{Ahn09}:
\begin{equation}
J_{21}(z)=(1+z)^{3}\int_{0}^{r_{{\rm LW}}}\frac{dr_{{\rm os}}}{1+z_{s}}\bar{j}_{\nu_{s}}(z_{s})f_{{\rm mod}}\left(r_{{\rm os}}\right),\label{eq:JLW}
\end{equation}
where $\bar{j}_{\nu_{s}}(z_{s})$ is the band-averaged emission
coefficient (in ${\rm erg}\,{\rm s}^{-1}\,{\rm Hz}^{-1}\,{\rm
  sr}^{-1}\,{\rm cm}^{-3}$) that is calculated from the
  emissivities generated by both Pop III and metal-enriched stars in
the rest-frame photon energy range {[}11.5,~13.6{]}~eV at the source
redshift z$_{s}$,
\begin{equation} 
r_{{\rm os}}\equiv2cH_{0}^{-1}\Omega_{m}^{-1/2}\left[(1+z_{{\rm obs}})^{-1/2}-(1+z_{s})^{-1/2}\right]
\end{equation}
is the comoving distance traveled by a photon from the source to
the observing point, and the intensity is modulated by the ``picket-fence
modulation factor'' $f_{{\rm mod}}$ given by
\begin{equation}
f_{{\rm mod}}=\begin{cases}
1.7\exp\left[-\left(\frac{r_{{\rm os}}/{\rm Mpc}}{116.29\alpha}\right)^{0.68}\right]-0.7 & {\rm if}\, r_{{\rm os}}\le97.39\alpha\,{\rm Mpc}\\
0 & {\rm otherwise},
\end{cases}\label{eq:fmod}
\end{equation}
where a scaling factor, 
\begin{equation}
\alpha=\left(\frac{h}{0.7}\right)^{-1}\left(\frac{\Omega_{m}}{0.27}\right)^{-0.5}\left(\frac{1+z_{{\rm s}}}{21}\right)^{-0.5},\label{eq:scaling_LW}
\end{equation}
defines the LW horizon $r_{{\rm LW}}\equiv97.39\alpha\,{\rm Mpc}$,
beyond which no sources contribute to $J_{{\rm LW}}.$ We calculate
$\bar{j}_{\nu_{s}}(z_{s})$ by averaging the source luminosity inside
the box as
\begin{equation}
\label{eq:jnu}
\bar{j}_{\nu_{s}}(z_{s}) = \frac{1}{4\pi l^{3}}
\sum_{i}\bar{L}_{\nu_{s},\, i}(z_{s}),
\end{equation}
where $\bar{L}_{\nu_{s},\, i}\equiv\int_{11.5\,{\rm eV}}^{13.6\,{\rm
    eV}}dE L_{\nu_{s},\, i}/(2.1\,{\rm eV})$ is the band-averaged
luminosity \citep{Schaerer02} of source $i$ and $l$ is the (proper)
size of the simulation box.

Figure \ref{fig:LW_background} shows the intensity of the LW
background used in the simulations of Normal region and Rarepeak in
units of J$_{21}$ (10$^{-21}$ erg s$^{-1}$ cm$^{-2}$ Hz$^{-1}$
sr$^{-1}$).  The background in the Rarepeak simulation is computed
with Equation (\ref{eqn:lwb}) and is higher at early times because the
mass resolution is not fine enough to capture the smallest
star-forming halos with masses $\sim$$2 \times 10^5~\Ms$
\citep{Tegmark97, Machacek01}.  At these very high redshifts, Pop III
stellar radiation dominates the emissivity, and by excluding the
low-mass end, we underestimate the background intensity in the Normal
region as seen in Figure \ref{fig:LW_background}.  Nevertheless, the
LW background is strong enough to delay the formation of Pop III stars
in low mass halos ($<$ 10$^7$ M$_\odot$)~\citep{OShea08}, but
eventually will be dominated by local sources with active stars.


The Void simulation runs the fastest because of the small amount of
structure formation relative to the other two simulations, and we are
able to evolve it down to $z=8$.  From the latest Planck results
\citep{Planck2015}, the universe is roughly half ionized at this
point.  Because we use the Normal simulation for the self-consistent
LW background, which ends at $z=12.5$, we do not use a LW background
for the Void simulation and only consider LW radiation from internal
sources.  Thus, we expect higher Pop III star formation rates that
occur in smaller halos and an earlier transition to metal-enriched
stars.  We have re-run the Void simulation with the \citet{Wise05} LW
background prescription and found that the impact is modest,
increasing the total Pop III stellar mass in the simulation by 7\% and
does not affect the metal-enriched star formation rate density
\citep[see Figure 2 in][]{Xu16_P3}.  We will not discuss the detailed
effects of the LW background on first galaxy formation and cosmic
evolution in this paper, though our simulations have shown that the
different LW background has some important impacts on star formations
at high redshift and in low mass halos. They do not significantly
change the results of this paper on ionizing photon production and
their escape fractions.

\begin{table*}
\centering
\caption{Summary of simulations 
\label{tab1}}
{
  \begin{tabular*}{0.99\textwidth}{@{\extracolsep{\fill}} ccccccccccc}
    \hline
    \multicolumn{3}{}{}  & \multicolumn{3}{c}{N$_{\rm halo}$} & & & &
    \multicolumn{2}{c}{f$_{\rm H}$} \\
    \cline{4-6} \cline{10-11} 
    \noalign{\vskip 1pt} 
    Region   & z & $\delta \rho$ & $>10^7 M_\odot$ & $>10^8 M_\odot$ &
    $> 10^9 M_\odot$ & N$_{\rm III}$ & M$_{\rm II}$ (M$_\odot$) &
     N$_{\rm ion}$               & vw & mw   \\ 
    (1)      & (2) & (3)  & (4) & (5) & (6)  & (7) & (8) & (9) & (10) & (11) \\  \hline
    \noalign{\vskip 1pt}
    Void     & 8        & -0.256       & 3263     & 172             & 5                         & 5110          & 6.27 $\times$ 10$^8$     & 3.87 $\times$ 10$^{69}$ & 0.117     & 0.132            \\ 
    Normal   & 12.5     & 0.0927        & 3275     & 137             & 3                         & 6544          & 1.92 $\times$ 10$^8$     & 1.20 $\times$ 10$^{69}$ & 0.020     & 0.030            \\ 
    Rarepeak & 15       & 0.686        & 3675               & 174     & 4                         & 10112         & 6.09 $\times$ 10$^8$     & 2.98 $\times$ 10$^{69}$ & 0.056     & 0.076            \\ \hline
\end{tabular*}

\parbox[t]{0.99\textwidth}{\textit{Notes:} Column (1): Name of
  simulation. Column (2): Redshift. Column (3): Overdensity of the
  region. Columns (4) to (6): Number of halos with mass larger than
  10$^7$, 10$^8$, and 10$^9$ M$_\odot$, respectively. Column (7):
  Number of Pop III stars and remnants. Column (8): Metal-enriched
  stellar mass. Column (9): Total number of ionizing photons generated
  during the simulation. Columns (10) and (11): Volume-weighted and
  mass-weighted hydrogen ionization fraction.}

}
\end{table*}


\subsection{Caveats and model dependencies}



Although the Renaissance Simulations include most of the relevant
physical processes during early galaxy formation, there are still some
missing physics and model dependencies, not dissimilar to other first
galaxy simulations \citep[e.g.][]{Muratov13b, Jeon14, Ricotti16}.
Most of these shortcomings arise from subgrid modeling of star
formation and feedback with the remaining dependencies originating
from large-scale and limited resolution effects.

The shape and characteristic mass of the Pop III initial mass function
(IMF) is weakly constrained, but it is most probably top-heavy
\citep[e.g.][]{ABN02, Susa14, Hirano15}.  The specific luminosity,
production of metals and stellar remnants, and their multiplicity are
all dependent on the IMF and the details of metal-free star formation
\citep{Schaerer02, Heger03}.  For instance if the characteristic mass
shifts from 20~\Ms{} to 60~\Ms{}, this could favor black hole
formation instead of neutron star formation and metal enrichment.
Furthermore, we do not consider X-ray radiative feedback from these
stellar remnants because of the computational expense of transporting
optically-thin X-rays, which could alter the thermal and ionization
properties of the ISM \citep{Alvarez09, Jeon13}, but it is not clear
whether it would have an impact on the galactic properties.  Massive
($10^4 - 10^6~\Ms$) black hole seeding and its feedback
\citep{Aykutalp14} is also neglected, however this process is most
likely rare \citep[e.g.][]{Dijkstra08, Agarwal16} and only impacts
very few high redshift galaxies.

Metal-enriched star formation is modeled on the scale of stellar
clusters with a minimum particle mass of $10^3~\Ms$.  Here we assume a
Salpeter IMF, which could not necessarily hold at high redshifts
\citep{Smith09, Safranek16}.  The effects of binary stellar evolution
and runaway stars, both of which can boost the UV escape fraction on
the $\sim$5\% level \citep{Conroy12, Kimm14, Ma16}, are not accounted
for in the simulation.  On the same topic, we take the ionizing
luminosity to be the average over the 20 Myr lifetime of the star
particle, whereas stellar population synthesis models show that the
luminosities are drop precipitously after 4 Myr.  Thus, we
underestimate the effects of radiative feedback, in particular from
radiation pressure \citep{Wise12b}, in the first few Myrs of each star
particle and overestimate the effects in the last $\sim$10 Myr.

On the large scale, relative streaming velocities ($v_{\rm rel} \sim
30 \unit{km} \unit{s}^{-1}$ at $z \sim 1100$) between baryons and dark
matter that arise during recombination \citep{Tseliakhovich10} can
suppress Pop III star formation in the smallest minihalos with $M_{\rm
  vir} \la 10^6~\Ms$ \citep[e.g.][]{Greif11, Naoz12, OLeary12}.
Albeit for computational reasons, our limited mass resolution
suppresses any cooling and star formation in halos with $M_{\rm vir}
\la 3 \times 10^6~\Ms$, roughly mimicking the same effect.  Although
we try to model the LW radiation background self-consistently, it is
not applied to all of the simulations, given the nature of zoom-in
simulations.  This affects Pop III star formation on the 10\% level
\citep{Xu16_P3}.  Lastly because zoom-in simulations do not capture
star formation outside of the focus region, any radiative feedback,
i.e. photo-suppression of low-mass galaxies
\citep[e.g.][]{Efstathiou92, ThoulWeinberg96}, from such ``external''
sources are not included and may become important when \hii~regions
start to overlap in the latter stages of reionization.


\section{Simulated Galaxy Properties}
\label{sec:prop}
We focus on the role of the first generations of galaxies in cosmic
reionization.  We first explore their global properties in the three
different regions of the {\it Renaissance Simulations}.  We then
calculate the distribution of various baryonic properties of the
star-forming halos and the ensuing reionization from this galaxy
population.  We then describe the UV luminosity function of these
galaxies and discuss the similarities of our simulated galaxies in
different survey volumes at various times.  In the our analysis, we
only include metal-enriched stars for the following reasons.  Because
massive Pop III stars are short-lived, it is rare for them to exist in
metal-enriched galaxies after a halo merger.  Additionally lower mass
($< 8~\Ms$) Pop III stars that may exist in such galaxies have a
negligible contribution to the stellar mass and luminosity because of
their low star formation efficiencies \citep[e.g.][]{Susa14} when
compared with the first galaxies \citep[e.g.][]{Wise14}.

\subsection{Early Galaxy Properties in Different Environments}

\begin{figure}[t]
\begin{center}
  \includegraphics[width=1.0\columnwidth]{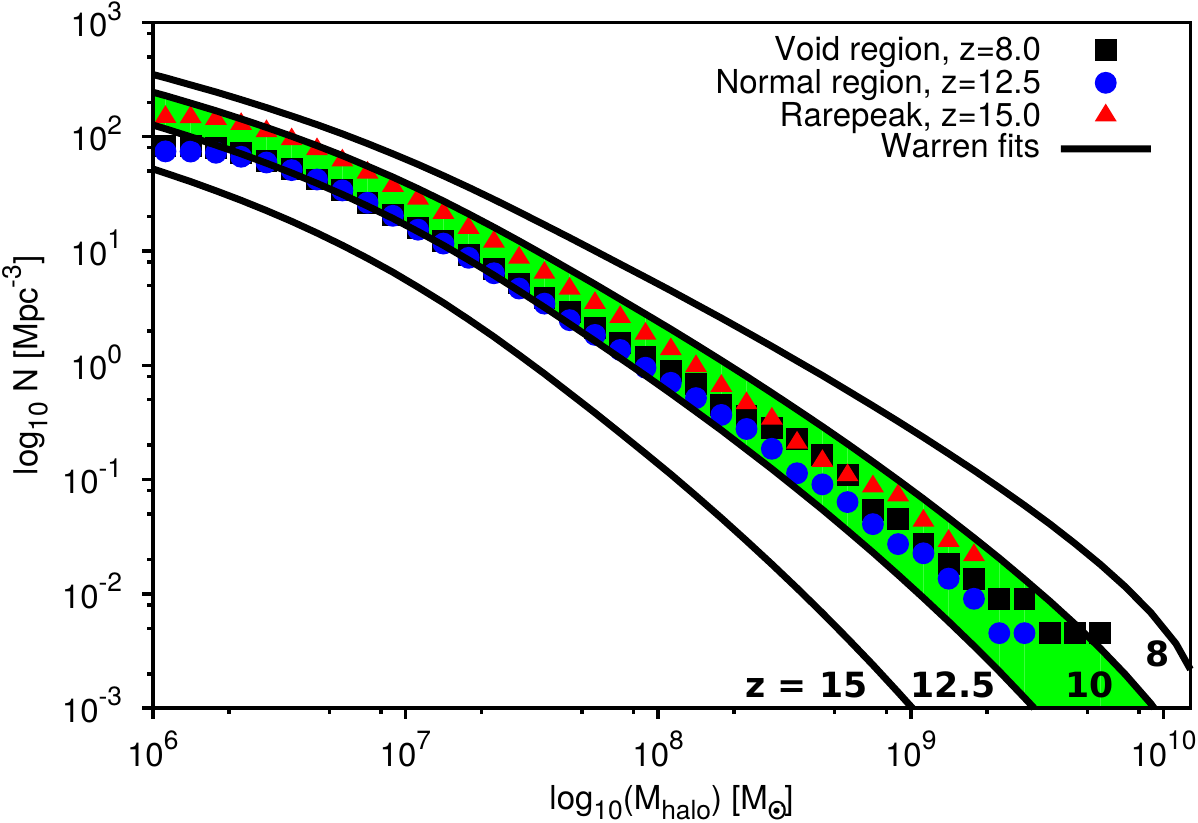}
\end{center}
\caption{Halo mass function within refined regions from the
  simulations (points) and the analytic fit (lines) from
  \citet{Warren06} at $z = 15, 12.5, 10,$ and 8.  The green shaded area
  denotes the halo number density from the \citeauthor{Warren06} fit
  between $z=12.5$ and $z=10$, and the simulated halo mass function of
  all of the regions lie within this range. \label{fig:HMF}}
\end{figure}

\begin{figure}[t]
\centering
\includegraphics[width=1.0\columnwidth]{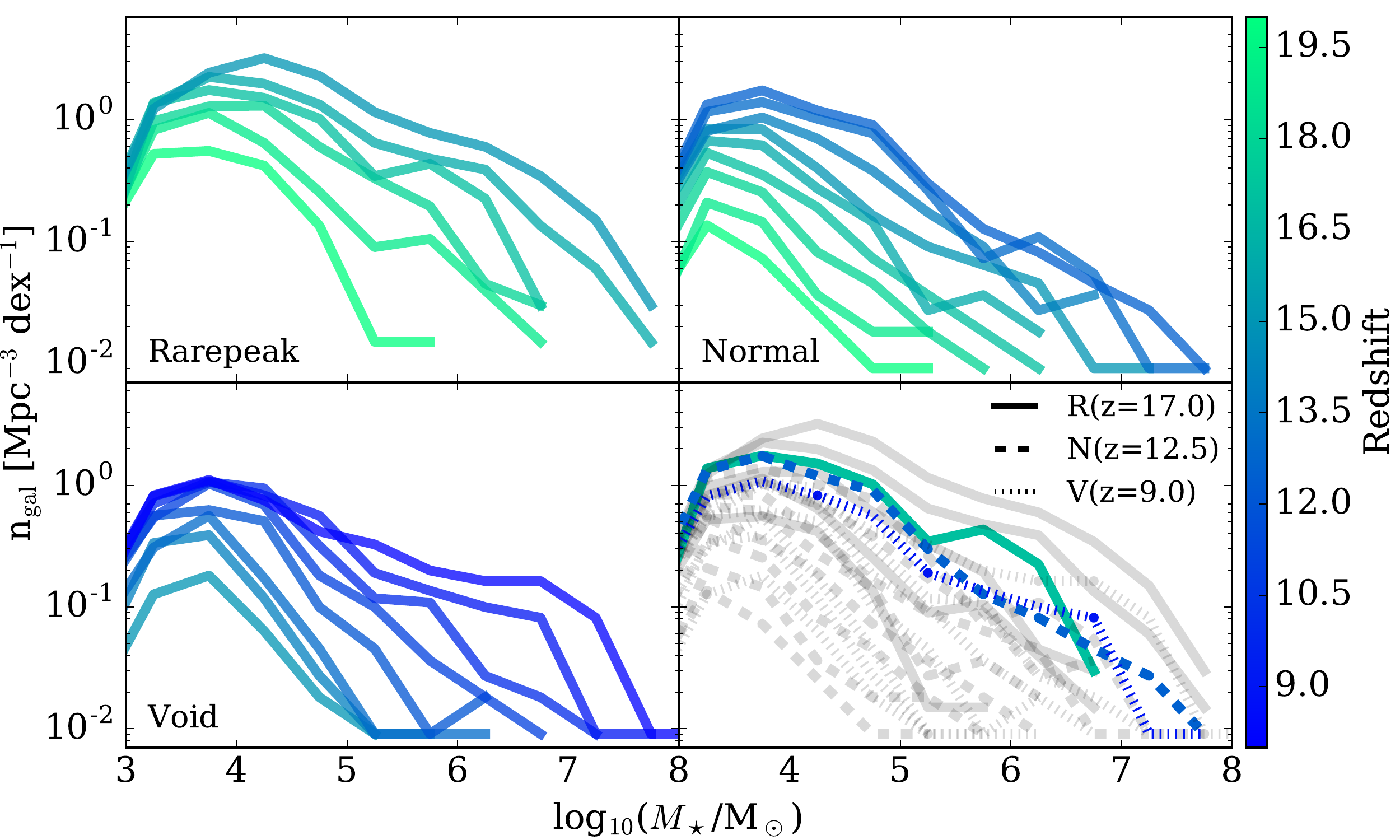}
 \caption{Stellar mass functions of the Rarepeak (top left), Normal
   (top right), and Void (bottom left) regions at redshift intervals
   $\Delta z = 1$ between redshifts 15--20, 13--20 (including 12.5),
   and 8--15, respectively.  The lower right panel shows the stellar
   mass functions from all of the regions and redshifts. The colored
   lines denote the redshifts (shown in the legend) that the most
   similar between the regions.
   \label{fig:smf}
 }
\end{figure}

We first present the global properties of the three survey volumes at
the final simulation redshift $z = (15, 12.5, 8)$ for the Rarepeak, Normal, and
Void regions, respectively.  Table 1 summarizes the overall
overdensity, halo counts, Pop III star counts, metal-enriched stellar
masses, total number of ionizing photons produced, and ionized
fraction.  Each simulation has more than 3,000 halos with masses $M >
10^7 \Ms$ that are viable hosts for star formation during the epoch of
reionization.  Between 5,000 and 10,000 Pop III stars form in each of
the regions, leading to over $10^8~\Ms$ of metal-enriched star
formation after the initial enrichment.  In total, over $10^{69}$ UV
ionizing photons are emitted from these stellar populations that
ionize 13\%, 3\%, and 8\% of the mass in the Void, Normal, and
Rarepeak volumes at the final redshift.

Before interpreting the results from the simulations, we first check
how representative these regions are of a typical patch of the
universe.  By construction, the Rarepeak, Normal, and Void simulations
capture high, mean, and low density regions, respectively.  Figure
\ref{fig:HMF} shows the simulated halo mass functions (HMFs) compared
to the analytic mass function of \citet{Warren06}, which is
calibrated by various $N$-body simulations and ellipsoidal collapse
models of Press-Schetcher formalism \citep{PS74, Sheth99}.  All of the
HMFs reside between the Warren fits at $z = 12.5$ and 10, depicted by
the shaded region in Figure \ref{fig:HMF}.  The Normal HMF at $z =
12.5$ fits well with the analytic fit at $z=12.5$ between $5 \times
10^6$ and $10^8~\Ms$, but the volume contains some small-scale
overdense modes, resulting in an overabundance of halos with $M >
10^8~\Ms$.  The Rarepeak and Void HMFs have much higher and lower halo
number densities than the analytic fit, demonstrating that they
truly vary from the cosmic mean.  The Void HMF is greater than the
Warren fit at $z=12.5$ in the high-mass end that is caused by a few
overdense smaller scale regions, similar to the Normal region.

The abundance of galaxies is inherently connected to the HMF.  Figure
\ref{fig:smf} shows the evolution of the stellar mass function (SMF)
in each of the regions, illustrating the initial assembly of galaxies,
starting at $z = 20$ in the Rarepeak and Normal regions and at $z =
15$ in the Void region.  As time progresses, more massive galaxies
form through {\it in-situ} star formation and mergers all while
low-mass galaxies continue to form.  Only in the Void region, the
abundance of low-mass galaxies with $M_\star \la 10^{4.5}~\Ms$ stop
increasing, suggesting that they are suppressed from either radiative
or stellar feedback, which we will investigate more closely later in
the section.  At higher stellar masses, the SMFs decrease with mass,
as expected.  However at any given redshift, the amplitudes in each
region vary because of the differences in the underlying HMF.  We
compare all of the simulated SMFs in the lower-right panel of Figure
\ref{fig:smf} to demonstrate the different mean assembly histories of
galaxies.  However, at some redshift, the SMFs will be similar between
regions.  Using the SMF at $z=12.5$ in the Normal region as a basis,
we find that the SMF in the Rarepeak and Void regions are the most
similar at $z = 17$ and 9, respectively.  The most striking difference
in the Rarepeak is the overabundance at the high-mass end caused by
the rapid assembly of the most massive galaxy.  In the Void region,
low-mass galaxies are suppressed by various forms of stellar feedback.

\begin{figure}[t]
\begin{center}
  \includegraphics[width=1.0\columnwidth]{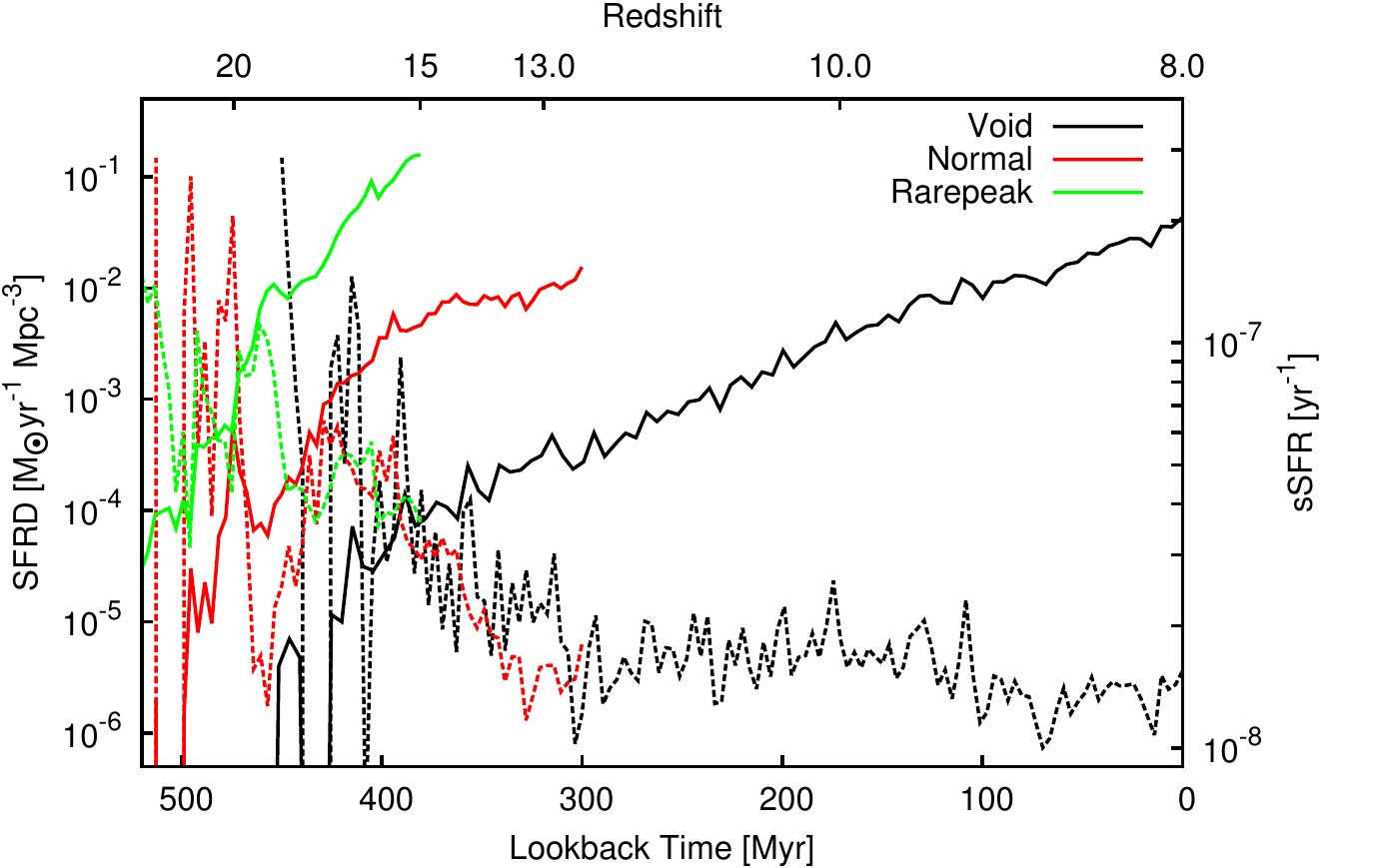}
\end{center}
\caption{Evolution of star formation rate density (solid lines) and
  specific star formation rate (dashed lines).  The star formation
  densities increase more rapidly in the more dense regions, as expected.
\label{fig:sfr}}
\end{figure}

We show the star formation rate densities (SFRD) and specific star
formation rate, (sSFR = SFR/M$_\star$) for all three regions in Figure
\ref{fig:sfr}.  In each region, the different large-scale
overdensities drive SFRD variations with the Rarepeak (Void) region
forming stars over an order of magnitude higher (lower) than the
Normal region at any given time.

\begin{figure}[t]
  \centering
  \includegraphics[width=1.0\columnwidth]{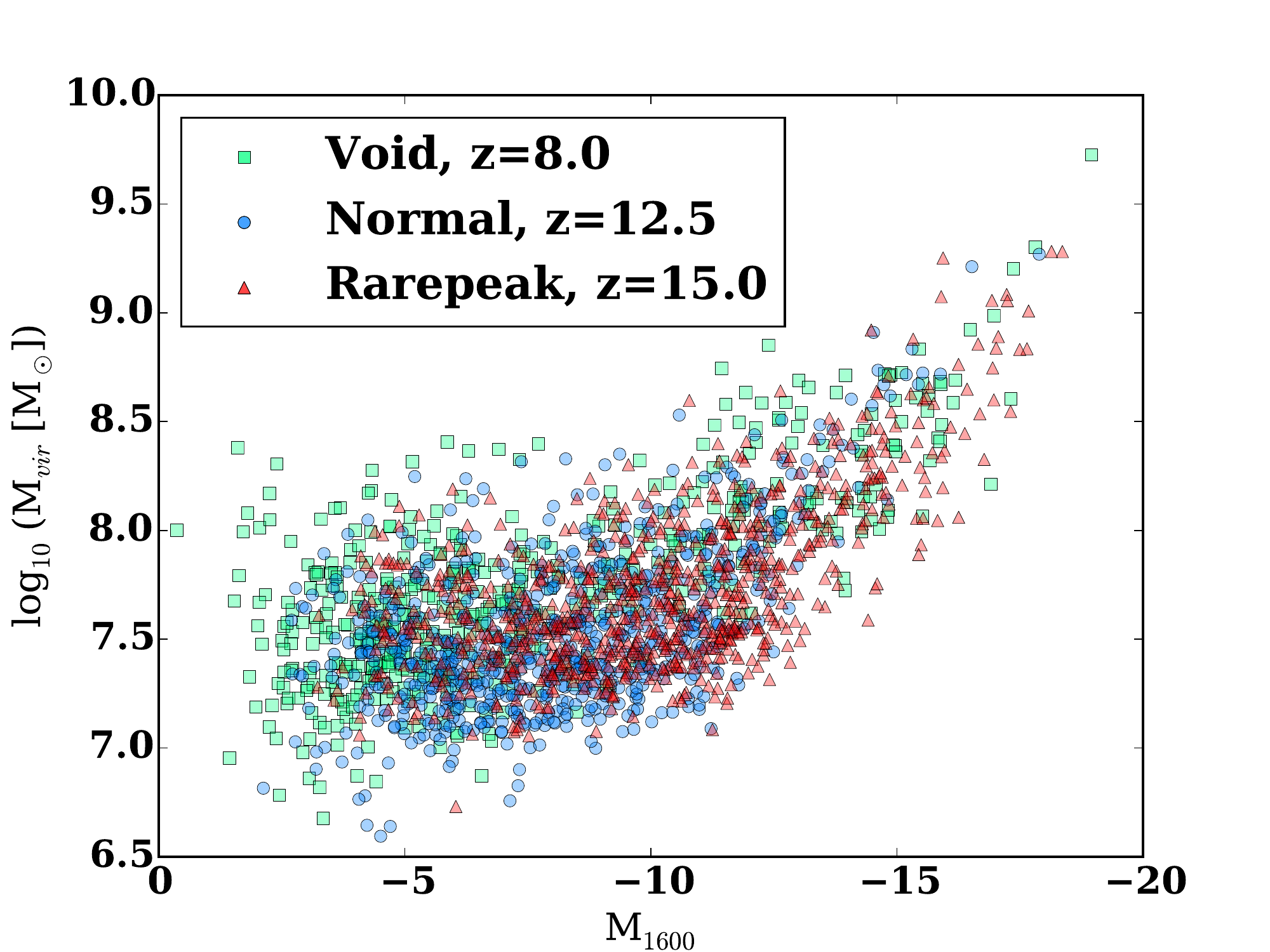}
  \hfill
  \includegraphics[width=1.0\columnwidth]{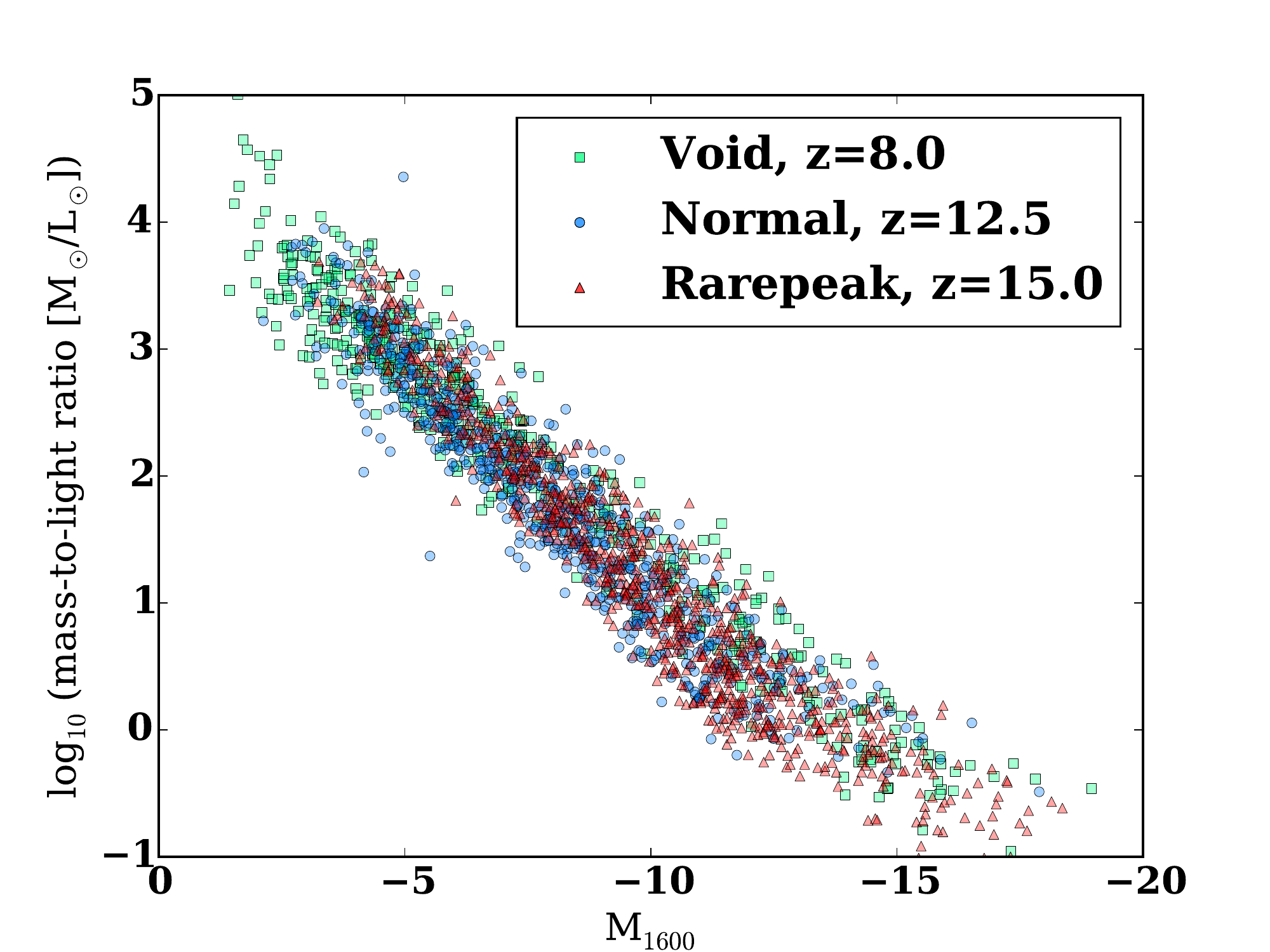}
  \caption{Scatter plots of virial mass (top panel) and mass (virial) to light
    ratio (bottom panel), in solar units,  as functions of UV magnitude
    for the Void region at $z=8$ (green squares), the Normal region at $z=12.5$ (blue circles)
    and the Rarepeak at $z=15$ (red triangles), respectively. Each
    point represents a single galaxy.
    \label{fig:MLratio}}
\end{figure}

\begin{figure}[t]
\centering
\includegraphics[width=1.0\columnwidth]{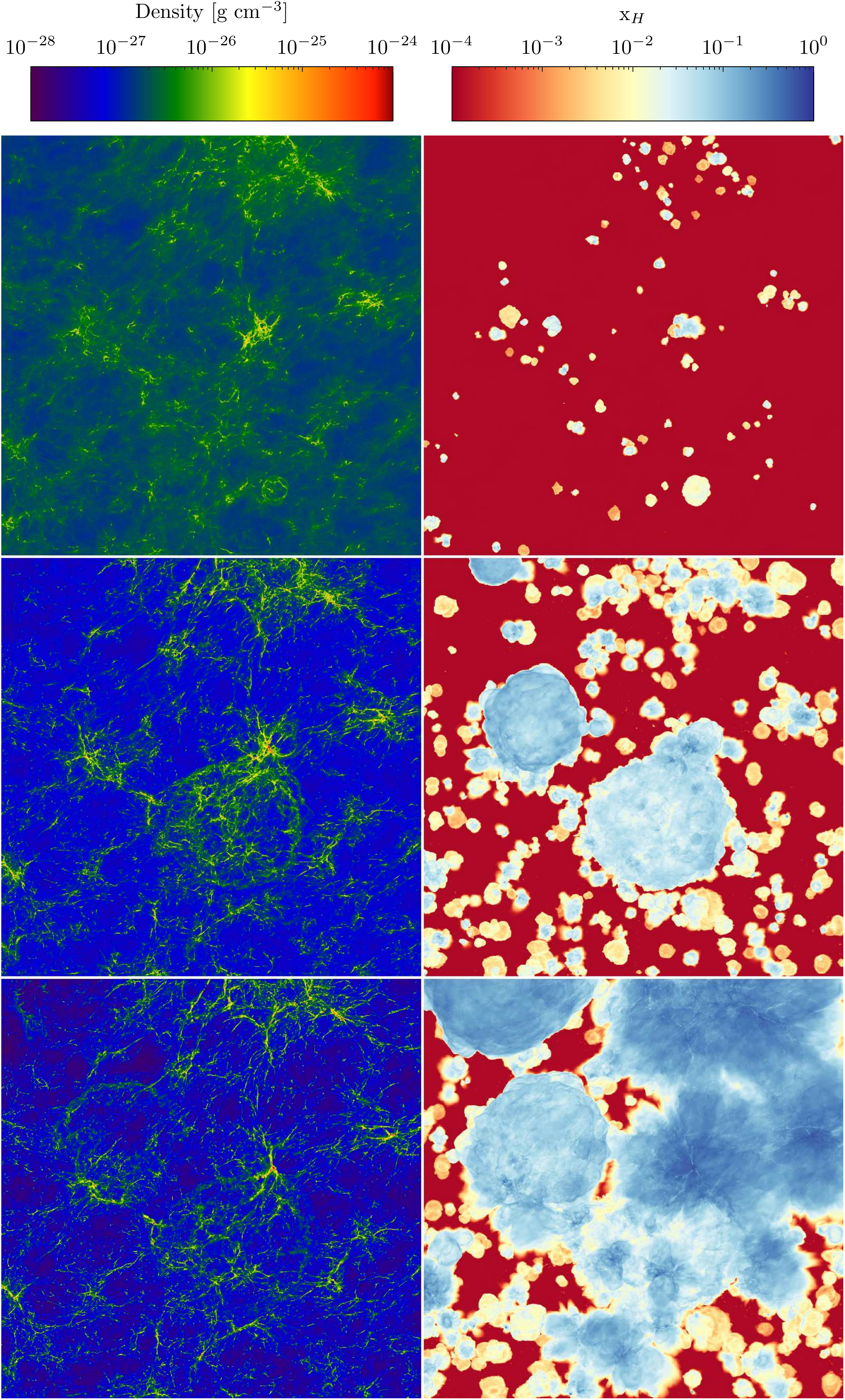}
\caption{Projections of density-weighted baryon density (left) and hydrogen
ionization fraction (right) of the Void region at $z=15$ (top), 10 (middle) and
8 (bottom). The projected volume is a cube with sides of 6.1 comoving Mpc.
\label{fig:slice_void}}
\end{figure}

\begin{figure}[t]
\centering
\includegraphics[width=1.0\columnwidth]{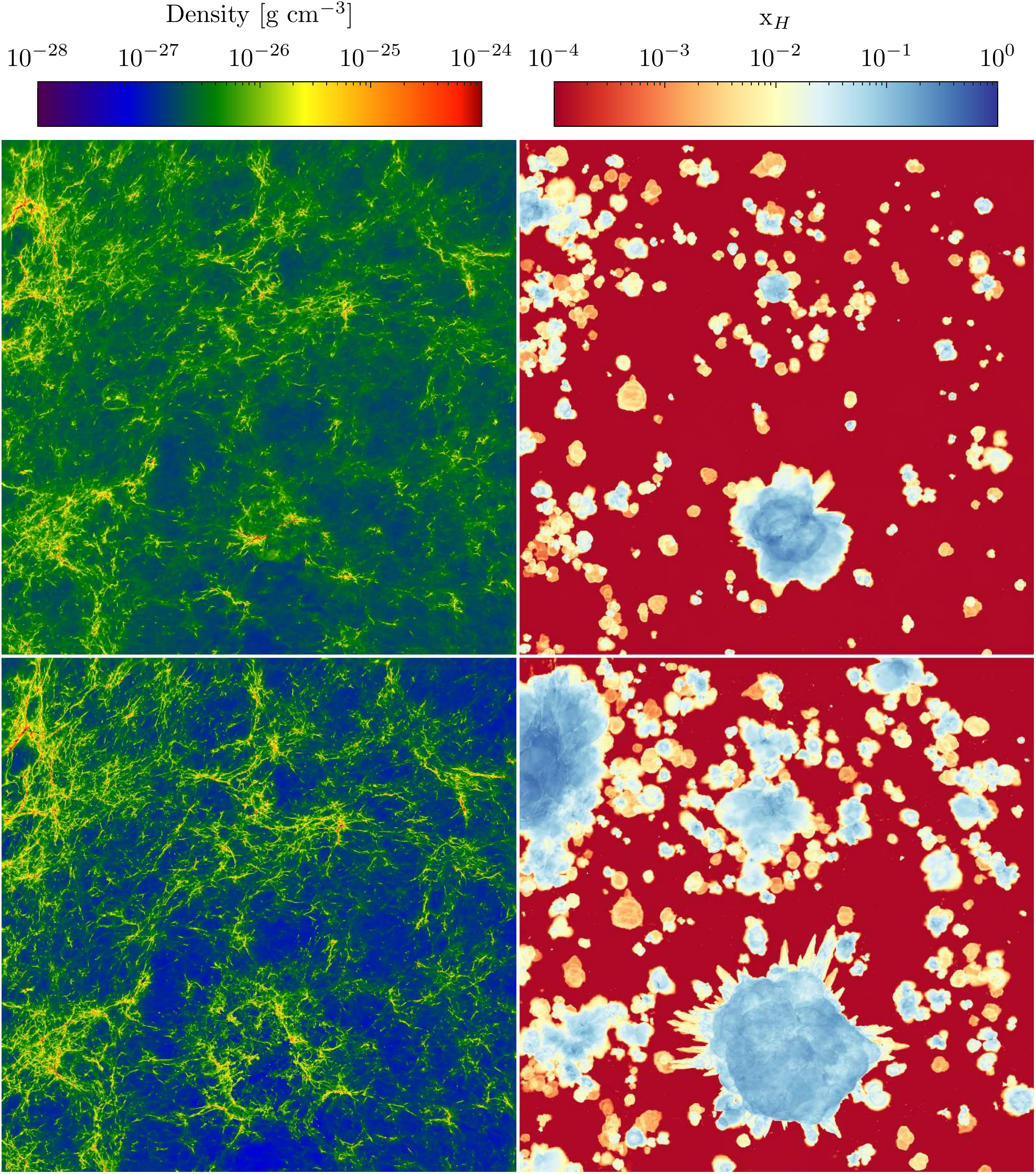}
\caption{Same as Figure \ref{fig:slice_void} but for the Normal region
  at $z=15$ (top) and $z=12.5$ (bottom).
\label{fig:slice_normal}}
\end{figure}

\begin{figure}[t]
\centering
\includegraphics[width=1.0\columnwidth]{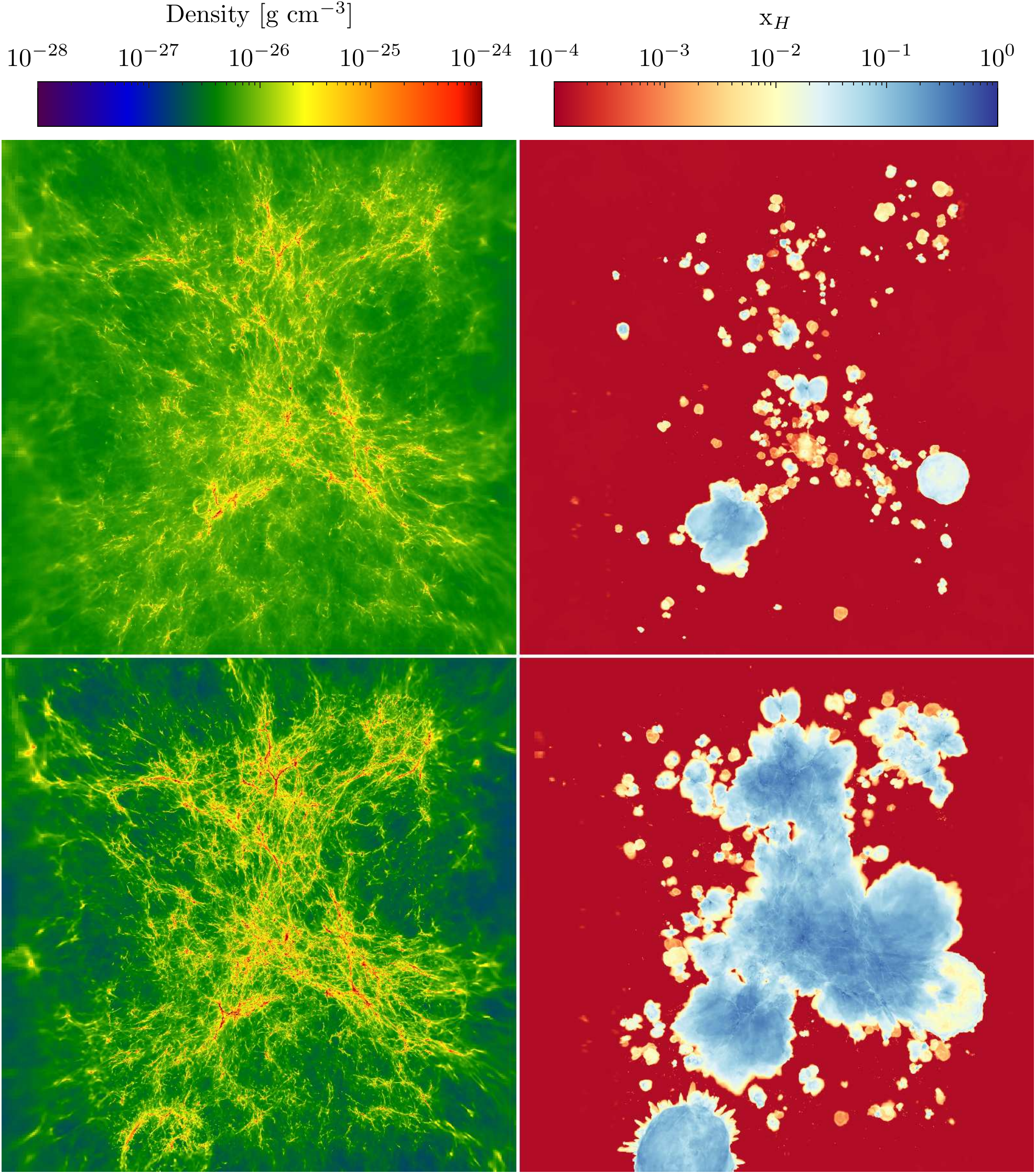}
\caption{Same as Figure \ref{fig:slice_void} but for the Rarepeak region
  at $z=18.5$ (top) and $z=15$ (bottom).
\label{fig:slice_rarepeak}}
\end{figure}

The number density of halos and star formation histories of the three
simulated regions are quite different.  One important question to
raise about simulations that probe different environments is whether
these galaxies can be considered to be a single population that mainly
depends on halo mass without much variation on environmental factors
and redshift during the early stages of cosmic reionization.  Figure
\ref{fig:MLratio} shows the virial mass and mass-to-light ratio as a
function of their total AB magnitude at 1600~\AA, M$_{1600}$.  To
compute the magnitude, we determine the spectral energy distribution
(SED) for each galaxy with the stellar population synthesis model of
\citet{Bruzual03}.  We use the ages, masses, and metallicities of the
metal-enriched star particles as input, assuming an instantaneous
burst model.  We do not consider any nebular emission lines in the
SEDs.  In galaxies with $M_{1600} \la -12$, brighter galaxies are
clearly hosted in larger halos.  Dimmer galaxies, however, are hosted
in halos with masses ranging from $3 \times 10^6$ to $3 \times
10^8~\Ms$.  These small and dim galaxies are usually the result of one
burst of star formation that has subsequently aged.  No new star
formation occurred afterwards as the gas supply has been disrupted by
supernova and radiative feedback.  The mass-to-light ratio shows a
more monotonic decreasing trend with increasing luminosity, similar to
those found in local dwarf galaxies \citep{McConnachie12}.  These two
trends are apparent in each of the three regions, where the overlap of
the data points suggest that they can be considered as a single galaxy
population that is independent of large-scale environment and
redshift, given that it is before cosmic reionization.  Thus, we
analyze the three regions as one galaxy sample throughout the rest of
the paper.  We discuss this simplification further in
\S\ref{sec:regions} before presenting our results on the UV escape
fraction.

\subsection{Ionization of the IGM}

\begin{figure}[t]
\centering
\includegraphics[width=1.0\columnwidth]{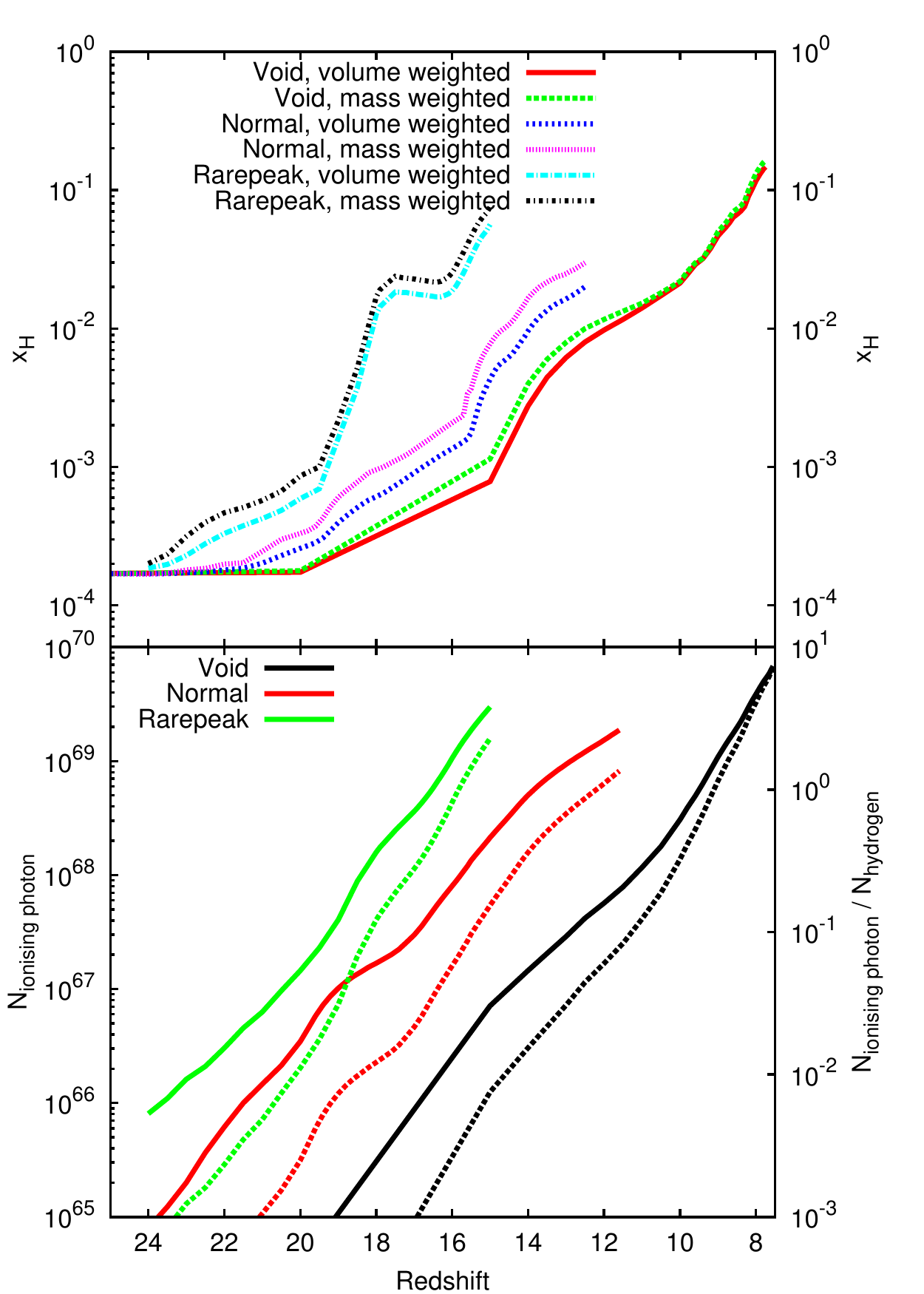}
\caption{Top panel: volume-weighted and mass-weighted ionization
  fractions as a function of redshift for all simulations. Bottom
  panel: cumulative number of ionizing photons (solid) from all
  galaxies and the ratios of number of ionizing photons to number of
  hydrogen atoms (dotted) within the survey
  volumes. \label{fig:Ionization_history}}
\end{figure}

We show projections of baryon density and hydrogen ionization fraction
of the Void, Normal, and Rarepeak regions in Figures
\ref{fig:slice_void}--\ref{fig:slice_rarepeak}, respectively, that use
the same color scales at various redshifts.  The differences in the
large-scale overdensities are clearly seen in the density projections.
The standard reionization picture is apparent in the Void simulation,
starting with isolated \hii~regions at higher redshifts (top row;
$z=15$).  These then grow and merge, forming progressively larger
volumes at later times, resulting in several $\sim$cMpc scale
\hii~regions at $z=8$.  In the process, any clumpiness in the IGM is
diminished as they are photo-heated \citep[e.g.][]{Pawlik09}. In the
Void simulation, a significant fraction of the ionizing radiation
leaves the survey volume, and they are not considered in the
calculation of the ionization fraction.  Interestingly, the galaxies
in the Rarepeak produce a similar amount of ionizing photons by $z=15$
as the Void region at $z=8$ (see Figure \ref{fig:Ionization_history}).
The typical \hii~regions are much smaller, however; this is caused by
higher recombination rates, which are proportional to $n^2$ in ionized
regions and thus strongly scale as $(1+z)^6$ that results in a
difference of a factor of $\sim$30 between the final redshifts of the
Rarepeak and Void regions.

Predictably, the ionization history in our three survey volumes differ
substantially.  Figure \ref{fig:Ionization_history} shows the
ionization fraction, total number of ionizing photons emitted, and the
ratio of ionizing photons to hydrogen atoms.  The Rarepeak region, as
expected, proceeds to form stars and ionize the region at a much
faster pace than the other regions.  Because the \hii~regions are
small and confined at $z \ga 12$ in all simulations, the mass-weighted
ionization fractions are significantly higher than the volume-weighted
ones.  For all simulations at their ending redshift, the
photon-to-baryon ratio is greater than unity, demonstrating that most
photons are lost to recombination processes over cosmic time.

\subsection{Galaxy properties and UV luminosity function}
\label{sec:props}

\begin{figure*}[t]
\centering
\includegraphics[width=1.0\textwidth]{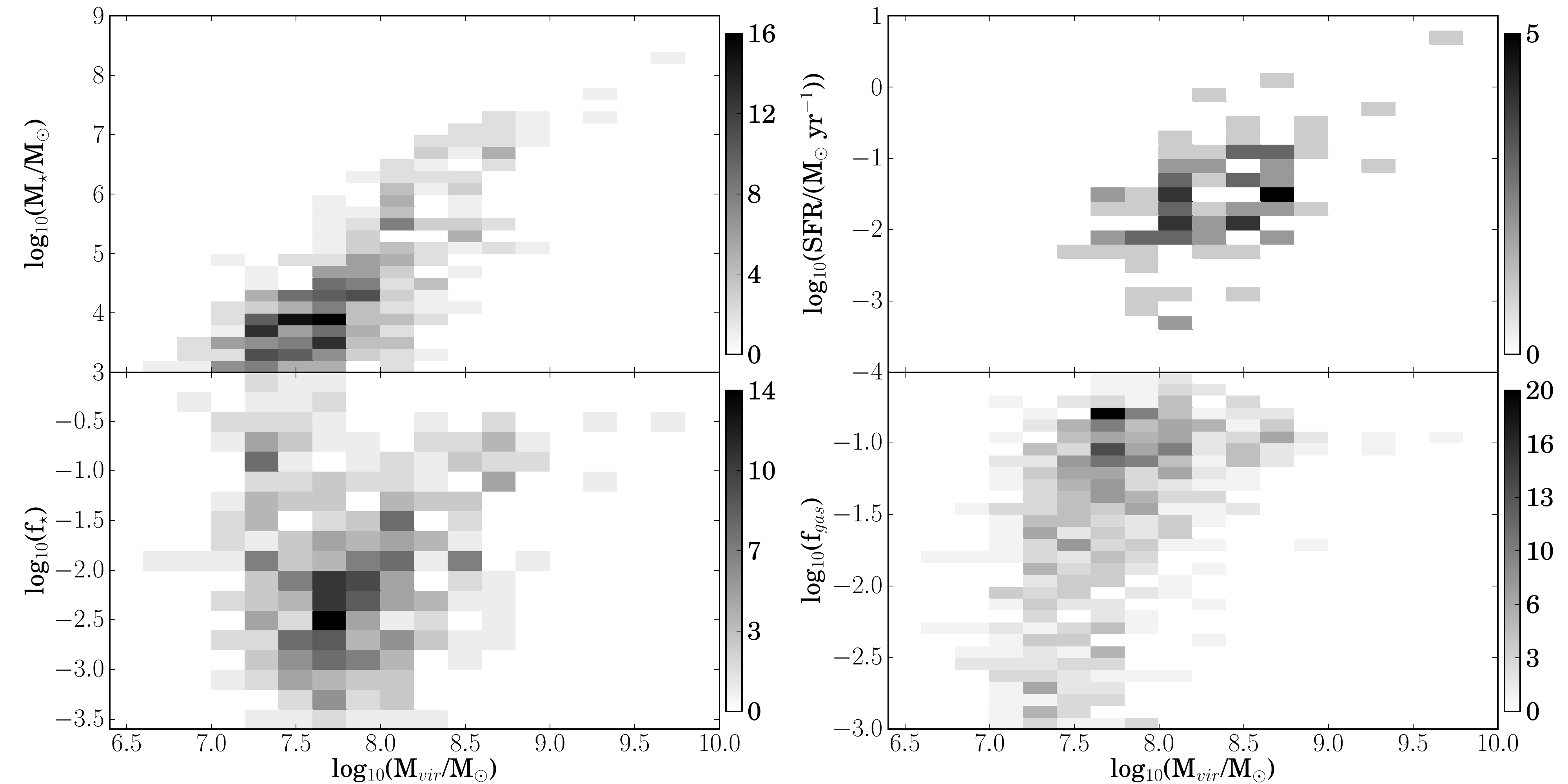}
\caption{ Distributions of (clockwise from the upper left) stellar
  mass, star formation rate, gas mass fraction and stellar baryonic
  fraction versus virial mass of all star-forming halos in the Void
  region at $z=8$. The pixels show the number of galaxies in the
  corresponding bins. \label{fig:star_virial_void}}
\end{figure*}

\begin{figure*}[t]
\centering
\includegraphics[width=1.0\textwidth]{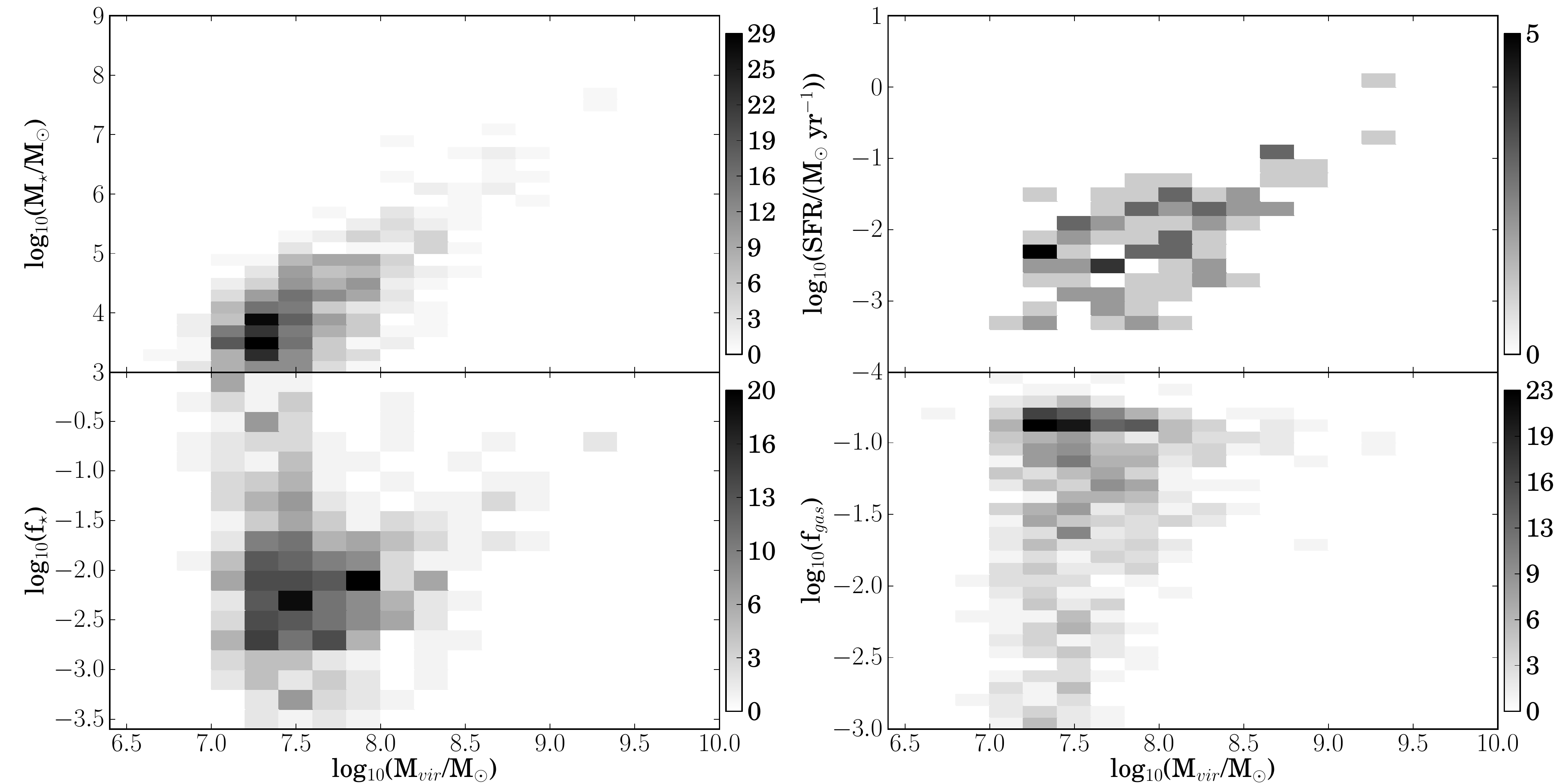}
\caption{Same as Figure \ref{fig:star_virial_void}, but for the normal region at $z=12.5$.
\label{fig:star_virial_normal}}
\end{figure*}

\begin{figure*}[t]
\centering
\includegraphics[width=1.0\textwidth]{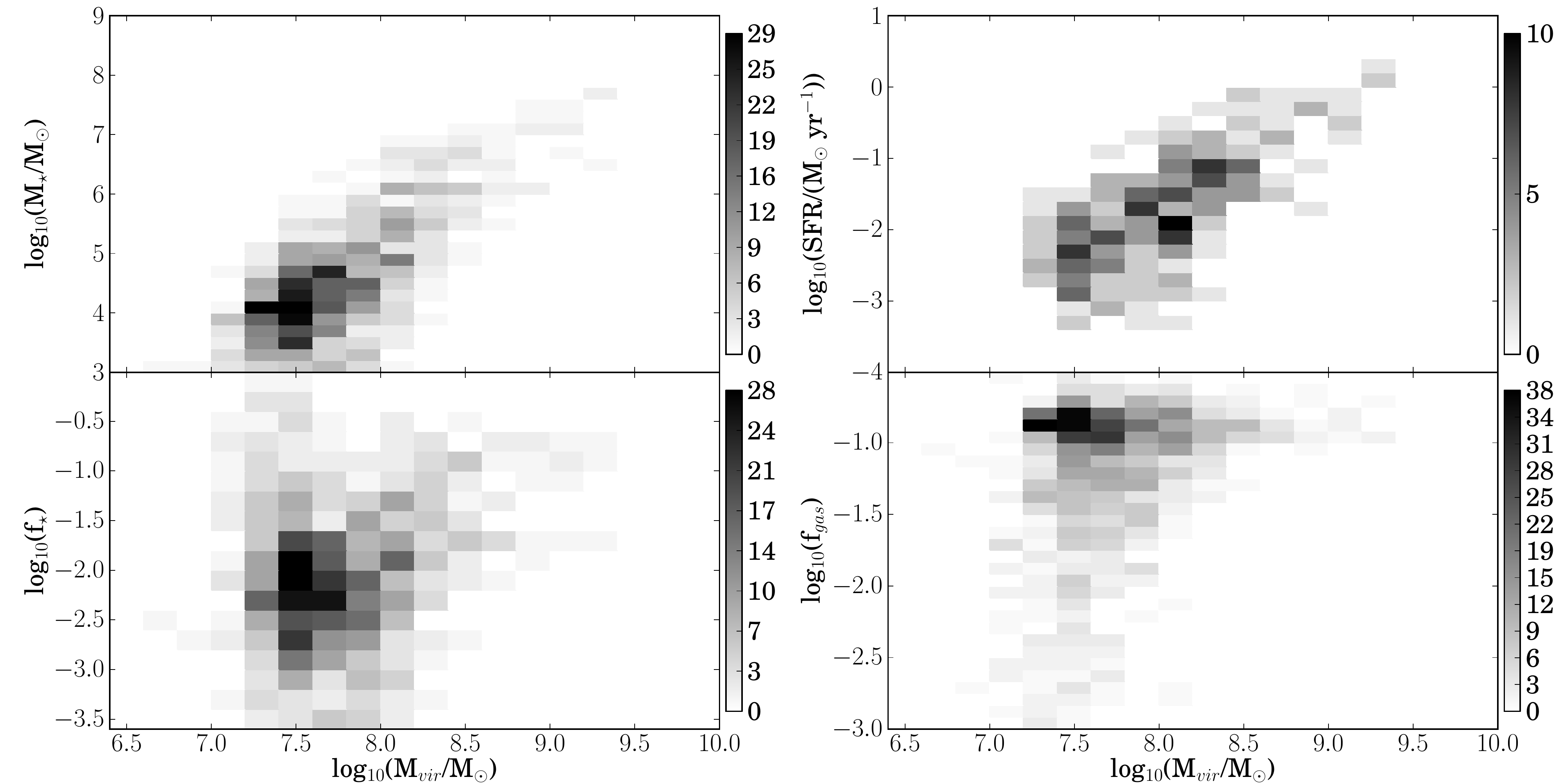}
\caption{Same as Figure \ref{fig:star_virial_void}, but for the Rarepeak at $z=15$.
\label{fig:star_virial_rarepeak}}
\end{figure*}

We now focus on the individual properties of the simulated galaxies in
our three survey regions.  We plot the distribution of the stellar
mass $M_\star$, star formation rate, gas mass fraction $f_{\rm gas}
\equiv M_{\rm gas}/M_{\rm vir}$, and stellar baryonic fraction
$f_\star \equiv M_\star / M_{\rm gas}$ as a function of halo mass
$M_{\rm vir}$ in Figures
\ref{fig:star_virial_void}--\ref{fig:star_virial_rarepeak} for the
Void, Normal, and Rarepeak volumes, respectively, at the final output
redshifts of $z = (8, 12.5, 15)$.  There are 468, 665, and 862 halos
that have hosted metal-enriched star formation in the Void, Normal,
and Rarepeak volumes, respectively, by these redshifts.  From a visual
inspection, the distributions from these regions look very similar,
suggesting that the galaxy properties are mainly determined by their
host halo masses.  Most of the halos with $10^7 \lsim M/\Ms \lsim
10^9$ have hosted metal-enriched star formation.  The largest halo in
the simulation suite is located in the Void region at $z=8$ with a
virial mass $M_{\rm vir} = 5.3 \times 10^9~\Ms$ and stellar mass
$M_\star = 1.7 \times 10^8~\Ms$.  There is very little star formation
in halos below a virial mass of $10^7~\Ms$, which is caused by a
combination of a strong UV incident radiation field, originating from
nearby galaxies, and the rapid mass accretion of the halos.  The UV
radiation can suppress star formation in low-mass halos at all
redshifts. However at earlier times, the UV radiation field is weaker,
allowing for star formation to occur in such halos, but by the final
redshift in the simulations most of these halos have merged into
larger galaxies.

The scaling relations shown in Figures
\ref{fig:star_virial_void}--\ref{fig:star_virial_rarepeak} are similar
to the ones found in \citet{Wise14}, who considered the same physical
processes and numerical methods as our work in a smaller (1 comoving
Mpc)$^3$ volume with a mass resolution that was 10 times smaller.  We
find the scatter in these relationships to be greater than in
\citet{Wise14} because our larger galaxy sample of $\sim$2000
galaxies, compared to their sample of 32 galaxies, probes various
large-scale environments and more importantly star formation
histories, where the gas properties are greatly affected by previous
stellar feedback in low-mass halos.  In each of our survey volumes,
the stellar mass to halo mass relation \citep[also see][for the
Rarepeak]{Chen14} is consistent with the extrapolated fitting formula
of \citet{Behroozi13}.  The gas mass fraction varies dramatically
between the extremes of nearly zero to the cosmic mean fraction in
halos with $M_{\rm vir} = 10^7 - 10^8~\Ms$.  These low-mass halos are
heavily affected by internal and external stellar feedback, cycling
between star-forming and quiescence phases \citep[cf.][]{Wise12b,
  Wise14, Hopkins13}, and their merger histories \citep{Chen14}.  Here
the overpressurized \hii~regions drive $\sim$30~\kms{} shocks
\citep[e.g.][]{Franco90} that alone can expel a large fraction of gas
from the shallow potential wells of these low-mass halos.  The
\hii~regions are anisotropic because of the turbulent nature of the
ISM, where the ionization front can propagate faster in directions
with smaller overall column densities.  This creates a porous ISM that
is an important factor in how much ionizing radiation escapes from the
halo, which will be examined in \S\ref{sec:fesc}.

Subsequent supernova feedback only exacerbate the ISM porosity and the
magnitude of the outflows, creating extremely gas-poor halos, as seen
in the $f_{\rm gas}$ panels of Figures
\ref{fig:star_virial_void}--\ref{fig:star_virial_rarepeak}.  The large
spread is caused by halos caught in different stages of its gaseous
disruption, where gas-poor halos have experienced a star formation
event tens of Myr earlier, gas-rich halos have just formed stars or
have accreted gas after an earlier star formation event, and the
intermediate halos are in the process of being evacuated of its gas
during an event or recovering through accretion.  As halos grow, it
becomes more difficult for gas to be expelled from the galaxy, but the
radiative and supernova feedback aid in stirring turbulence and
regulating star formation.

For even larger halos ($M_{\rm vir} \ge 10^9~\Ms$), the scatter in the
scaling relations is interestingly smaller even though there are only
3--5 halos in this mass range in each region, suggesting that galaxies
are less susceptible to feedback than low-mass halos and evolve on
tighter scaling relations.  Their gas fractions (f$_{gas}$) are
$\sim$0.1, and their stellar baryonic fractions (f$_{\star}$) lie
within the approximate range 0.03--0.1.  Although we only have a
sample of 12 star-forming halos at these masses, we find that their
properties primarily depend on their host halo masses and are less
dependent on their formation histories and the environment.

It is worthwhile to compare how $f_{\rm gas}$ in low-mass halos
depends on the environment to the assumptions made in some
semi-analytical studies.  For example, \citet{Dayal14}, in estimating
the galaxy luminosity function (LF) from halo merger trees, assumes
that $f_{\rm gas} = 0$ after any star formation activity in halos in
the mass range $10^8 - 10^9~\Ms$.  This assumption is only
qualitatively consistent with our findings of the gas fraction being
diminished in minihalos through radiative and supernova feedback;
however, the majority of them have gas fractions above 0.05.  The
results from such studies may be affected when updating the models to
include episodic star formation in these smallest star-forming halos.

\begin{figure}[t]
\centering
 \includegraphics[width=1.0\columnwidth]{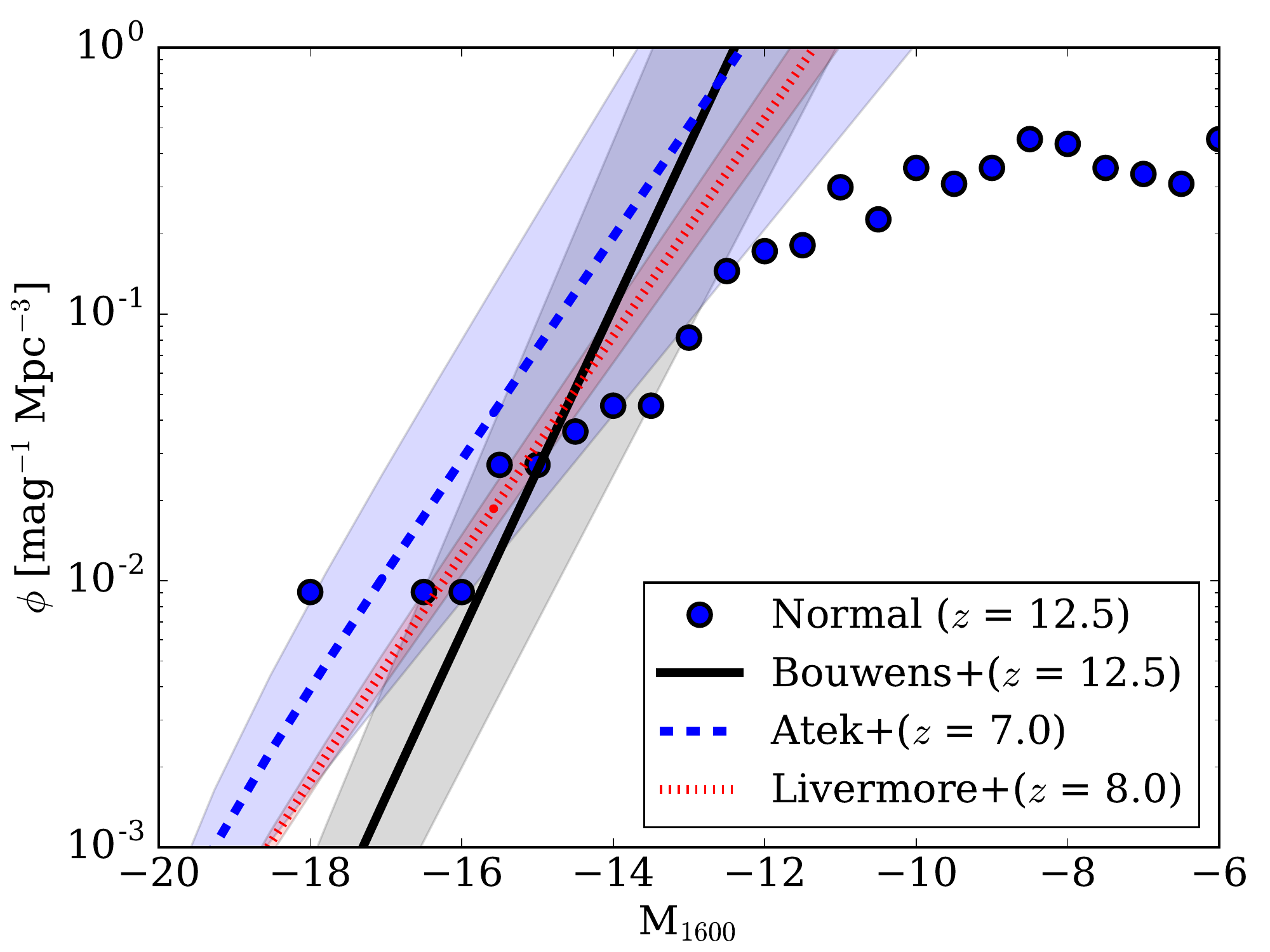}
 \caption{UV galaxy luminosity function for the Normal region at
   $z=12.5$ (circles) compared to observed luminosity functions in the
   Frontier Fields at $z=7$ \citep[dashed line;][]{Atek15} and $z=8$
   \citep[dotted line;][]{Livermore16} and in the HUDF, extrapolated
   to $z=12.5$ \citep[solid line;][]{Bouwens15}.  Because the slope
   and normalization have shown some dependence on redshift, the $z=7$
   and $z=8$ data should only be used as a reference point when
   interpreting the simulated luminosity function, which flattens from
   the observed faint-end power law at $M_{1600} \ga -12$.}
\end{figure}

\begin{figure*}
\centering
 \includegraphics[width=1.0\textwidth]{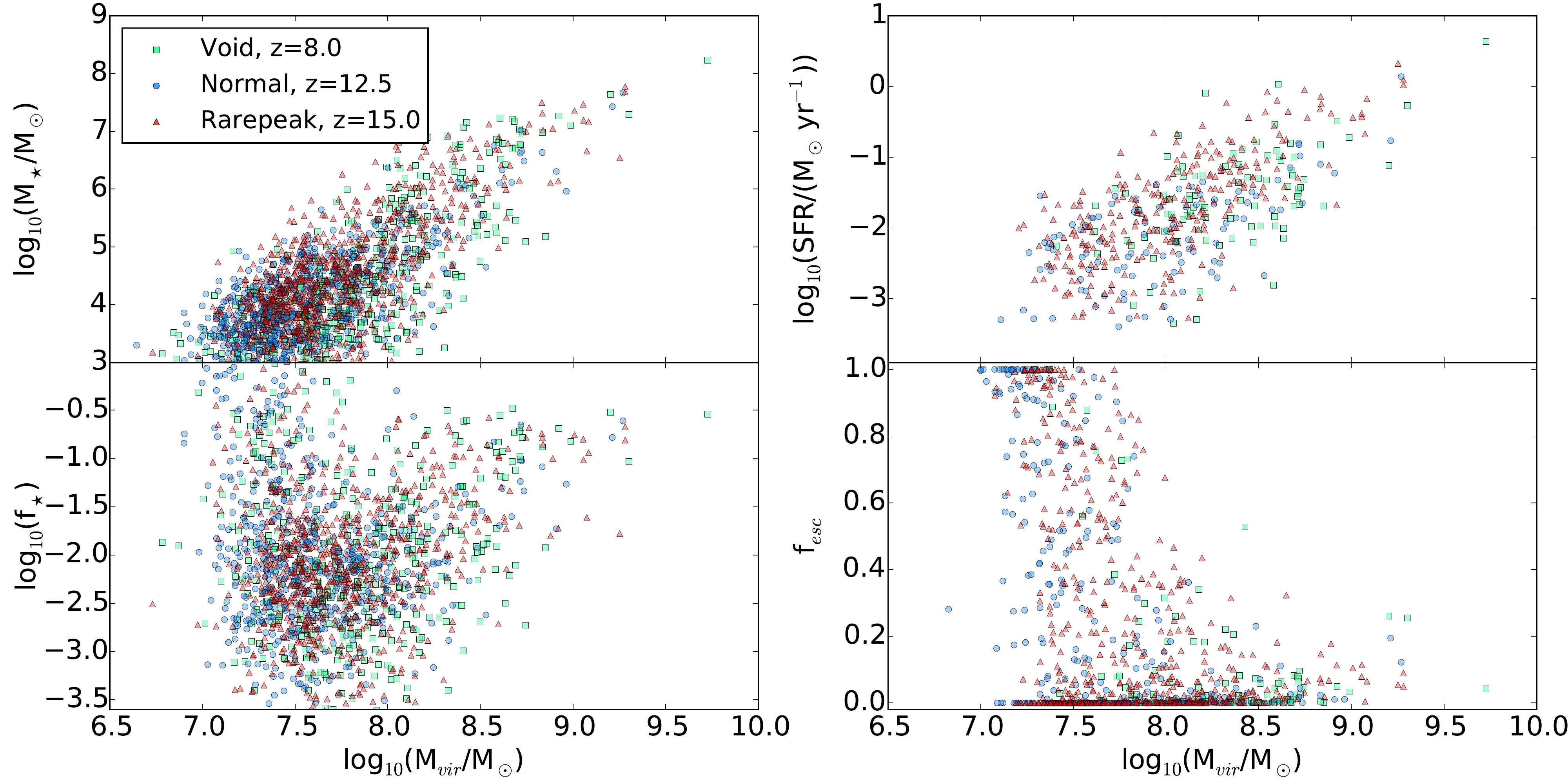}
 \caption{Distributions of (clockwise from the upper left) stellar
   mass, star formation rate, UV escape fraction, and stellar baryonic
   fraction versus virial mass of all star-forming haloes in all three
   simulations, Void region at $z=8$ (green square), Normal region at
   $z=12.5$ (blue circle) and Rarepeak at $z=15$ (red triangle). Each
   point represents a galaxy. \label{fig:galaxies_onefigure}}
\end{figure*}

\begin{figure}
\centering
 \includegraphics[width=1.0\columnwidth]{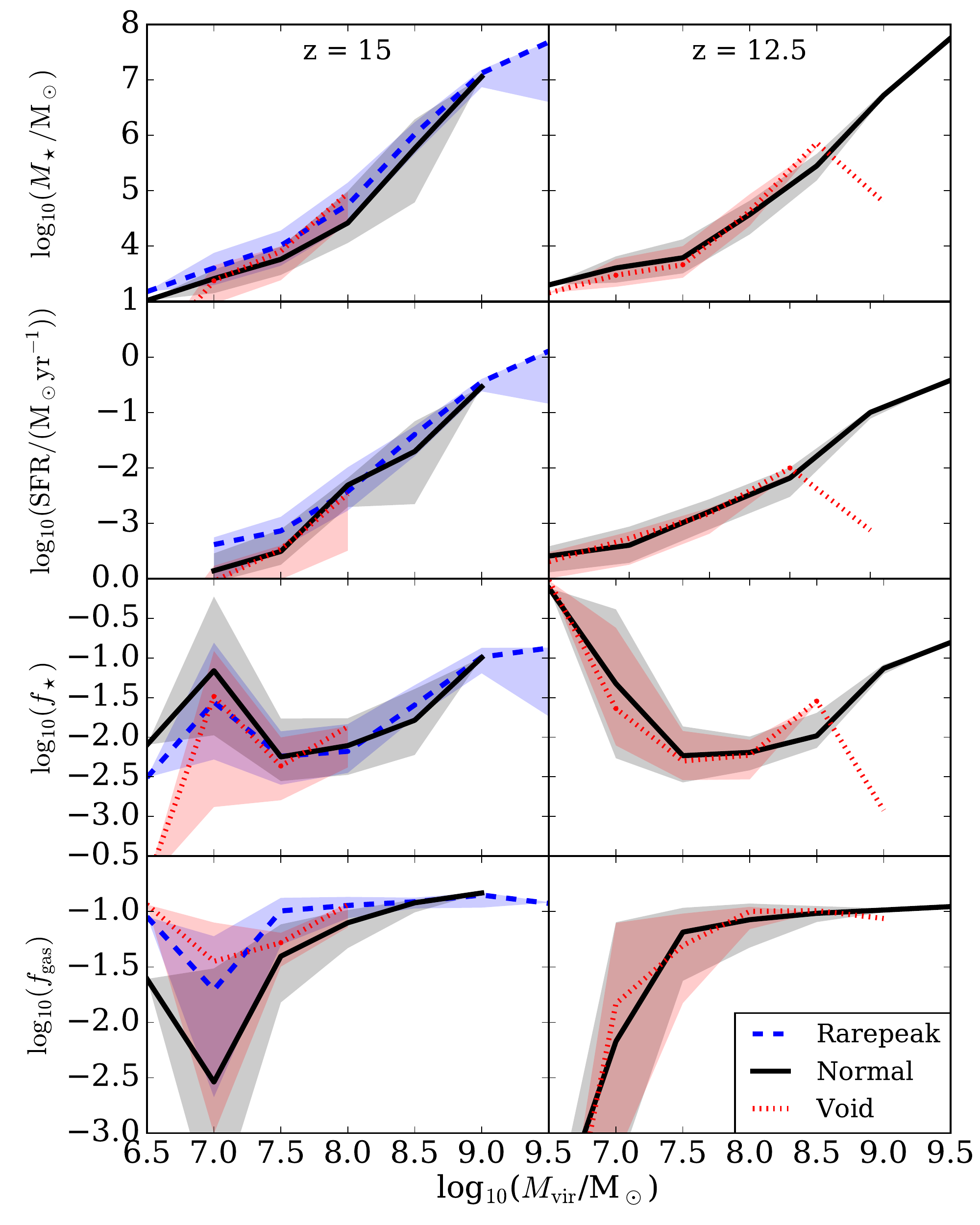}
 \caption{Comparison of (top to bottom) stellar mass,
   star formation rate, stellar baryonic fraction, and gas fraction trends
   with halo mass of the Rarepeak (blue dashed), Normal (solid black),
   and Void (red dotted) regions at redshift 15 (left) and 12.5
   (right).  The Rarepeak data do not exist for $z = 12.5$ because of
   its final redshift of 15.  The shaded regions denote 1-$\sigma$
   deviations in each halo mass bin.
   \label{fig:same_redshift}
 }
\end{figure}

The galaxy LF is an important quantification of the overall galaxy
population, especially when calculating the intrinsic galactic
ionizing emissivity.  Combined with the fraction of ionizing UV
radiation that escapes from the halo, the LF faint-end slope and the
behavior at very low luminosities determine the photons available for
cosmic reionization.  The aforementioned scaling relations or the UV
galaxy LF can be used in semi-analytic reionization calculations
\citep[e.g.][]{Robertson10, Robertson13, Robertson15, Alvarez12,
  Kuhlen12} to check whether galaxies can provide a sufficient amount
of radiation to complete and sustain reionization by $z \sim 6$.  We
have presented the evolution of galaxy LFs of the Renaissance
Simulations in \citet{OShea15}, but we reproduce the LF in the Normal
region at $z=12.5$, now comparing it to the latest results from the
{\it Frontier Fields} \citep{Atek15, Livermore16} and the
\citet{Bouwens15} redshift dependent fit.  The faint-end slope and
normalization should continue to steepen and decrease with redshift,
respectively, so the $z = 7$ and $8$ observed LFs should be used as a
reference when inspecting the simulated LF.  There are only a few
galaxies with M$_{\rm UV} < -17$ in our survey volumes, the limiting
(unlensed) magnitude of the HUDF at $z = 7-8$, and in this limit, our
simulated number densities are consistent with the LFs found in the
HUDF.  The simulated Renaissance Simulation LFs follow a power-law
with a slope consistent with the HUDF extrapolated to low luminosities
until they flatten at low luminosities of M$_{\rm UV} \gsim -12$.
This behavior is also seen in the higher resolution simulations of
\citet{Wise14}.

\subsection{Similarities of galaxies in various environments and
  redshifts}
\label{sec:regions}

\begin{table*}
\centering
\caption{Galaxy and host halo properties
\label{tab2}}
{
  \renewcommand{\arraystretch}{1.8}
  \begin{tabular*}{0.99\textwidth}{@{\extracolsep{\fill}}ccccccc}
    \hline
    \hline
 log M$_{\rm vir}$ [M$_\odot$]   & Region & log SFR [M$_\odot$ yr$^{-1}$]  & log f$_\star$  & f$_{\rm gas}$ & log M$_{\star}$ [M$_\odot$] &  f$_{\rm esc}$ \\[-0.75em]
   (1) & (2) & (3) & (4) & (5) & (6) & (7) \\[-0.25em]
\hline
\hline
\multirow{3}{*}{6.75}  & Void   & --  & -0.07$^{+0.98}_{-1.80}$ & 0.005$^{+0.004}_{-0.004}$   & 3.26$^{+0.22}_{-0.20}$  & --\\
\cline{2-7}
  & Normal & -- & 0.24$^{+1.24}_{-1.04}$  & 0.008$^{+0.007}_{-0.008}$  & 3.44$^{+0.33}_{-0.35}$  &   0.64$^{+0.24}_{-0.24}$\\
\cline{2-7}
  & Rarepeak & -- & -2.62$^{+0.07}_{-0.07}$  & 0.055$^{+0.030}_{-0.030}$  & 3.15$^{+0.02}_{-0.02}$ & -- \\
\hline
\hline
\multirow{3}{*}{7.25}  &  Void & --  & -1.26$^{+1.56}_{-1.39}$   & 0.016$^{+0.017}_{-0.016}$  &  3.56$^{+0.49}_{-0.46}$ & 0.58$^{+0.42}_{-0.57}$  \\ 
\cline{2-7}
  & Normal & -2.49$^{+0.38}_{-0.74}$  & -1.51$^{+1.94}_{-1.33}$ & 0.043$^{+0.081}_{-0.043}$  & 3.56$^{+0.58}_{-0.51}$  &  0.51$^{+0.49}_{-0.51}$  \\
\cline{2-7}
  & Rarepeak & -2.36$^{+0.39}_{-0.38}$  & -1.97$^{+1.31}_{-1.68}$  & 0.069$^{+0.076}_{-0.069}$  &  3.67$^{+0.71}_{-0.81}$ & 0.42$^{+0.55}_{-0.42}$  \\
\hline
\hline
\multirow{3}{*}{7.75}  &  Void & -2.09$^{+0.53}_{-0.32}$  & -2.23$^{+0.77}_{-0.80}$  & 0.058$^{+0.072}_{-0.057}$  & 4.02$^{+0.61}_{-0.63}$  & 0.12$^{+0.18}_{-0.12}$ \\
\cline{2-7}
  & Normal & -2.39$^{+0.62}_{-0.58}$  & -2.17$^{+0.52}_{-0.59}$  & 0.071$^{+0.072}_{-0.064}$  &  4.14$^{+0.71}_{-0.75}$ & 0.17$^{+0.31}_{-0.17}$  \\
\cline{2-7}
  & Rarepeak & -2.11$^{+0.57}_{-0.65}$  & -2.34$^{+0.68}_{-0.68}$  & 0.091$^{+0.065}_{-0.071}$  & 4.20$^{+0.74}_{-0.72}$ &  0.21$^{+0.35}_{-0.21}$  \\
\hline
\hline
\multirow{3}{*}{8.25}  &  Void & -1.65$^{+0.53}_{-0.53}$  & -2.02$^{+0.73}_{-0.82}$  & 0.102$^{+0.035}_{-0.029}$  & 5.11$^{+1.10}_{-1.20}$  & 0.04$^{+0.02}_{-0.04}$  \\
\cline{2-7}
  & Normal & -2.03$^{+0.50}_{-0.46}$  & -1.99$^{+0.46}_{-0.53}$  & 0.086$^{+0.046}_{-0.056}$  & 5.03$^{+0.63}_{-0.74}$  &  0.03$^{+0.01}_{-0.03}$ \\
\cline{2-7}
  & Rarepeak & -1.53$^{+0.57}_{-0.59}$  & -1.85$^{+0.68}_{-0.60}$  & 0.118$^{+0.048}_{-0.053}$  & 5.38$^{+0.78}_{-0.78}$ &   0.07$^{+0.09}_{-0.07}$  \\
\hline
\hline
\multirow{3}{*}{8.75}  &  Void & -1.33$^{+0.51}_{-0.58}$  & -1.33$^{+0.60}_{-0.66}$  & 0.104$^{-0.035}_{-0.029}$  &  6.32$^{+0.74}_{-0.99}$ & 0.03$^{+0.04}_{-0.03}$  \\
\cline{2-7}
  & Normal & -1.38$^{+0.47}_{-0.36}$  & -1.38$^{+0.37}_{-0.32}$  & 0.095$^{+0.038}_{-0.042}$  & 6.24$^{+0.42}_{-0.32}$  &  0.02$^{+0.02}_{-0.02}$ \\
\cline{2-7}
  & Rarepeak & -0.80$^{+0.57}_{-0.63}$  & -1.34$^{+0.56}_{-0.38}$  & 0.133$^{+0.021}_{-0.024}$  & 6.44$^{+0.61}_{-0.45}$ &  0.07$^{+0.05}_{-0.06}$  \\
\hline
\hline
\multirow{3}{*}{9.25}  &  Void & -0.69$^{+0.29}_{-0.29}$ & -0.78$^{+0.17}_{-0.17}$  & 0.097$^{+0.005}_{-0.005}$  & 7.46$^{+0.12}_{-0.12}$ & 0.26$^{+0.00}_{-0.00}$  \\
\cline{2-7}
  & Normal & -0.31$^{+0.31}_{-0.31}$  & -0.70$^{+0.06}_{-0.06}$  & 0.100$^{-0.001}_{-0.001}$  & 7.54$^{+0.81}_{-0.81}$  & 0.16$^{+0.02}_{-0.02}$  \\
\cline{2-7}
  & Rarepeak & -0.21$^{+0.29}_{-0.23}$  & -1.05$^{+0.28}_{-0.50}$  & 0.141$^{+0.021}_{-0.024}$  & 7.23$^{+0.43}_{-0.51}$  & 0.07$^{+0.03}_{-0.02}$ \\
\hline
\hline
\end{tabular*}
\parbox[t]{0.99\textwidth}{\textit{Notes:} Statistics are shown for galaxies of 
Void region at z = 8, Normal region at z = 12.5 and Rarepeak at z = 15 in 0.5 dex 
  bins in M$_{vir}$. Column (1): Center of mass bin. 
  Column (2): Simulations. Column (3): Star formation rate density. Dashes indicate no recent star formation. Column (4): 
  Stellar baryonic fraction. Column (5): Gas fraction. Column (6): Stellar mass. 
  Column (7): Fraction of hydrogen ionizing radiation that escape the virial radius.
  Errors shown are 1-$\sigma$ deviations.}
}
\renewcommand{\arraystretch}{1}
\end{table*}


We have shown that the general trends and scatter of various bulk
properties---UV luminosity, mass-to-light ratio, stellar masses, star
formation rates, stellar baryonic fractions, and gas fractions---
during the epoch of reionization are similar in each of the three
survey volumes.  The lack of environmental variation suggests that
galaxies during their initial formation are mainly dependent on their
host halo mass, and they are nearly independent of environment or
redshift, given that $z \gsim 8$.  

We now directly compare the stellar mass, stellar baryonic fraction,
star formation rate, and escape fraction (see \S\ref{sec:fesc}) as a
function of halo mass from each region in Figure
\ref{fig:galaxies_onefigure} at their final redshifts.  For a more
quantitative comparison, we also tabulate the galaxy properties and
escape fractions (see \S\ref{sec:fesc}) in 0.5 dex bins of M$_{\rm
  vir}$ in Table \ref{tab2}.  Although the galaxy number densities and
star formation densities vary greatly between the three survey
volumes, we see that the distribution of individual galaxy properties
are indistinguishable between large-scale environments with their
means and variations consistent with a single population. This
invariance suggests that environment and formation time play a lesser
role during the formation of the first galaxies.  In principle, these
effects can suppress or induce star formation through preheating,
enrichment, and/or halo temperatures \citep{Gnedin00}.

In halos with $M \gsim 10^8~\Ms$, we have found that the host halo
mass is the predominant factor, albeit with a large scatter, in
controlling galaxy formation during the epoch of reionization.  In
low-mass ($M \lsim 10^8~\Ms$) halos, the SFRs and stellar masses
increase with halo mass.  But the gas and stellar baryonic fractions
have a very large scatter, which is primarily caused by differing
histories of star formation and halo mass accretion.  Furthermore, the
scatter in these quantities is further magnified by capturing these
halos in different stages of stellar feedback, i.e. the evacuation of
gas.

Figure \ref{fig:same_redshift} compares the three regions at same
redshift ($z = 15, 12.5$) to explore how much environment affects
early galaxy growth.  At both redshifts, the median and variance of
the stellar mass and SFR are consistent in all regions, suggesting
that halo mass is a dominant factor in determining the stellar content
of halos.  We note that the SFR in the Void region trends lower than
the other regions.  At $z=15$, the stellar baryonic fraction and gas
fraction in the regions are within 1-$\sigma$ of each other with
greater variances at lower halo masses for reasons discussed
previously.  At $z=12.5$, the Normal and Void regions nearly mirror
each other in all four quantities shown in the Figure with the
exception of the highest mass bin that contains a peculiar halo that
has a very low stellar mass ($M_{\rm vir} = 6.0 \times 10^8~\Ms$,
$M_\star = 6.2 \times 10^4~\Ms$, $f_{\rm gas} = 0.087$) that warrants
a more detailed inspection in a later study.

By comparing our simulated galaxy properties in the Void, Normal, and
Rarepeak regions, we validate the method of \citet{Wise14} and
\citet{Chen14} to use galaxies from various redshifts in a single
simulation as a single sample to statistically study galaxy properties
and scaling relations.  We caution, however, that this technique
almost certainly breaks down at later times when most halos are
exposed to an external UV background, photoevaporating the low-mass
halos at the end and after the epoch of reionization.





\section{UV Escape Fraction and Photon Budget}
\label{sec:fesc}
For radiation sources in galaxies to contribute to cosmic
reionization, their photons must propagate into the nearby IGM, of
which only a fraction $f_{\rm esc}$ manage to escape.  The remaining
fraction are absorbed by neutral gas within the virial radius of the galaxy.  In order to
semi-analytically calculate the evolution of the ionized fraction, the
ionizing photon emissivity is required.  The rate of escaping ionizing
photons in a single halo can be parameterized as
\begin{equation}
  \dot{n}_{\rm \gamma, halo} = \psi_{\gamma} \, f_{\rm esc} \, f_{\star} \, f_{\rm
    gas} \, (\dot{M}_{\rm vir} / \mu m_{\rm H})
\end{equation}
where the product $f_{\star} f_{\rm gas} \dot{M}_{\rm vir}$ is the
halo's instantaneous SFR, $\mu$ is the mean molecular weight, and
$\psi_{\gamma}$ is the number of ionizing photons produced per stellar
baryon during a stellar lifetime, which can range from 6,000 for a
Salpeter IMF with solar metallicity to 13,000 for a metal-poor
($[\mathrm{Z/H}] = -3.3$) population star \citep{Schaerer03}.  The
escape fraction $f_{\rm esc}$ is the most uncertain factor in
determining the photon budget in reionization models and has attracted
much attention of numerical studies \citep[e.g.][]{Gnedin08_fesc,
  Wise09, Razoumov10, Yajima11, Paardekooper13, Ferrara13, Kimm14,
  Wise14, Ma15}.

\begin{figure}[t]
\centering
\includegraphics[width=1.0\columnwidth]{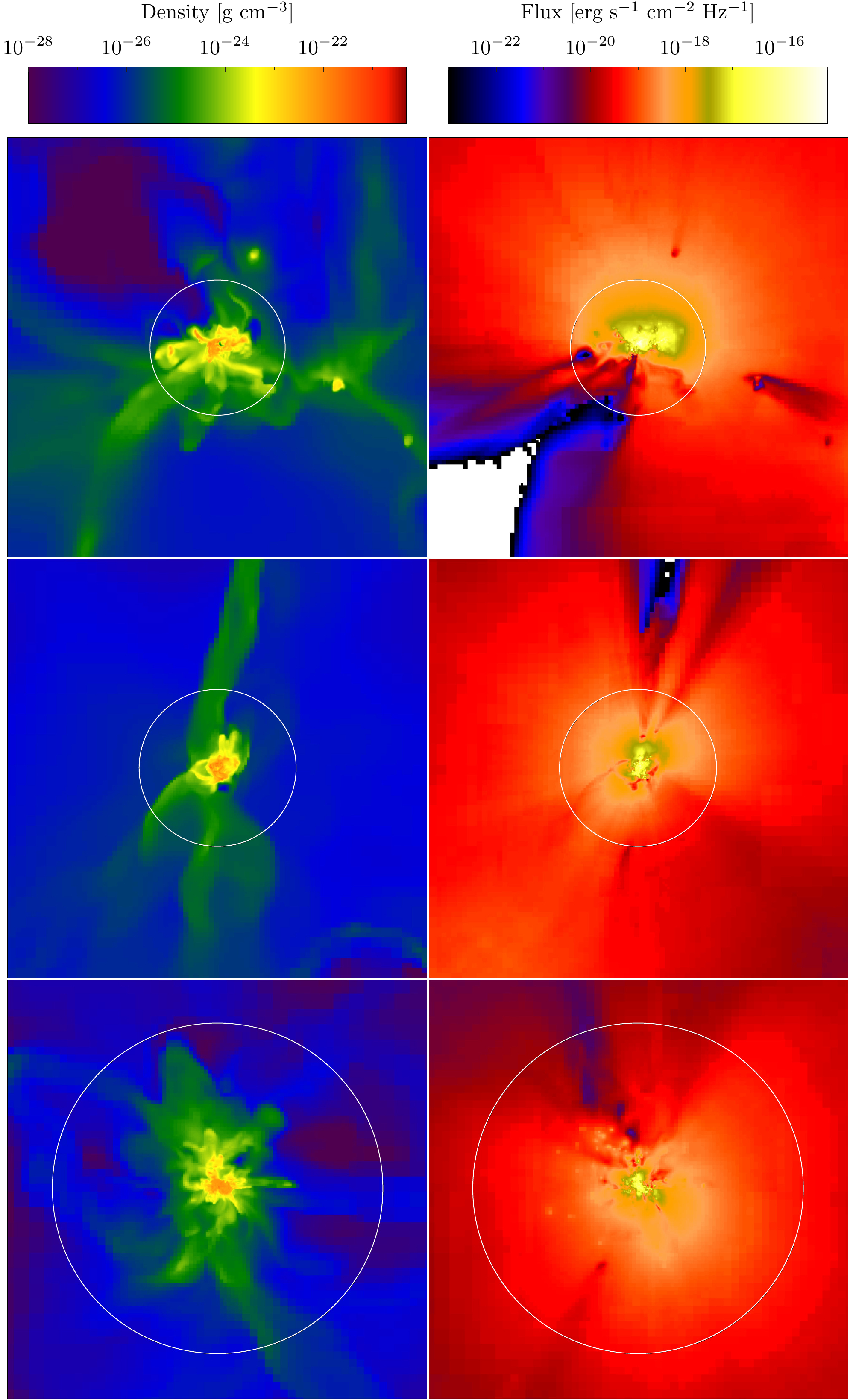}
\caption{Projections of density-weighted average baryon density (left)
  and UV ($E=21.6$ eV) flux (right) of the most massive halo in Void,
  Normal, and Rarepeak simulations at $z=8$, 12.5 and 15,
  respectively. These projections show the gaseous structure in the
  galaxy and the propagation of the UV photons. The projections have a
  field of view of 15 proper kpc and a depth of 600 proper pc. The
  white circles show the virial radii. \label{fig:halo_projects}}
\end{figure}

\begin{figure*}[t]
\centering
\includegraphics[width=1.0\textwidth]{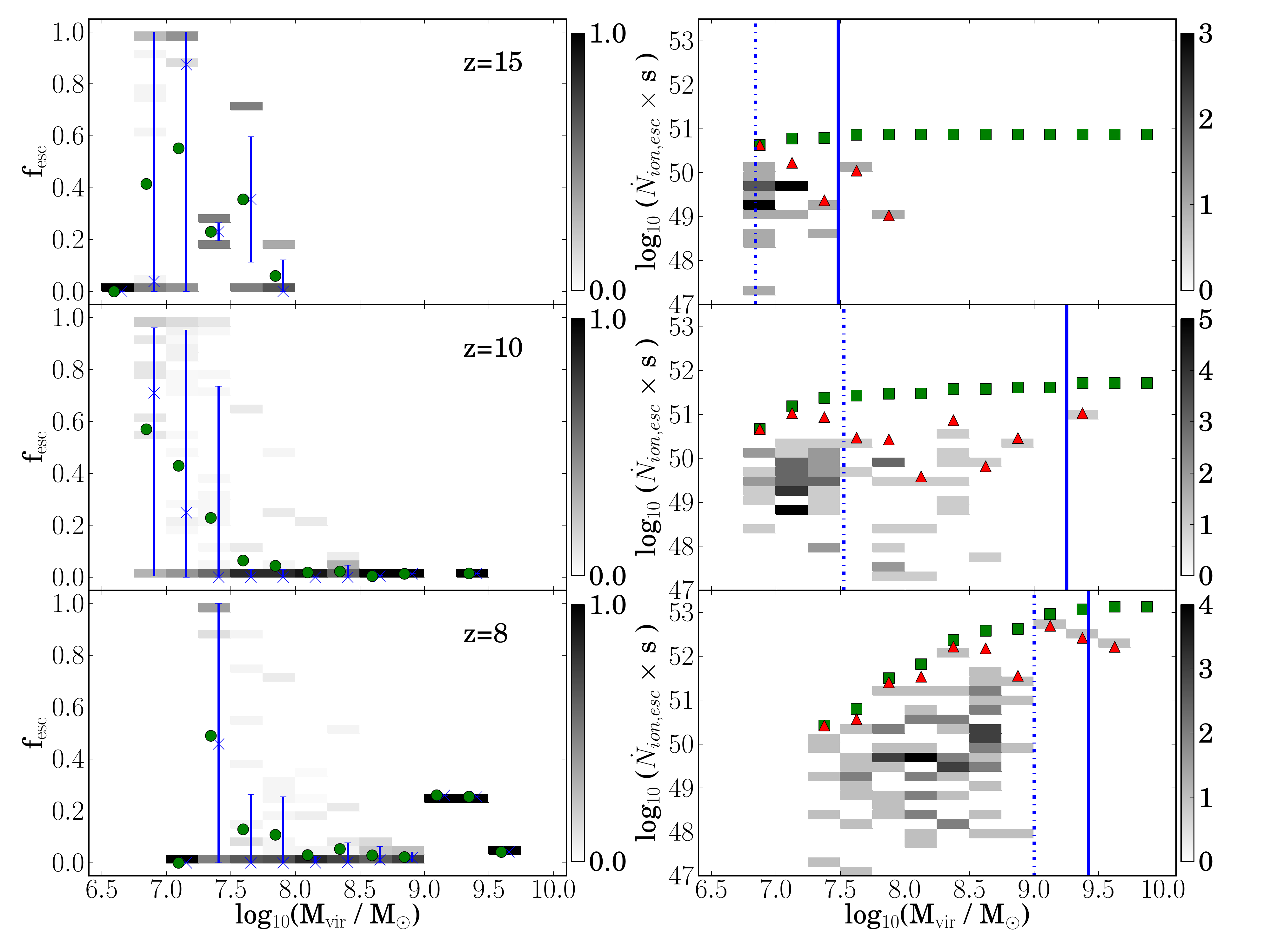}
\caption{ Probability density functions for the UV escape fraction
  f$_{\rm esc}$ (left) and number of escaped UV photons (right) as a
  function of halo virial mass at $z=15$ (top), 10 (middle) and 8
  (bottom) for Void region simulation.  The mean escape fractions in
  0.25 dex bins are represented by green filled circles. The median
  escape fractions are represented by blue crosses, and 15.9 and 84.1
  percentiles are shown as the bottom and top of vertical lines. The
  number of escaped UV photons from galaxies and the
  cumulative number of escaped UV photons in 0.25 dex bins are represented by red
  triangles and green squares, respectively. The vertical dash-dotted
  and solid lines show the mass bins at which the cumulative UV escaped
  photons are half and 90\% of the total escaped
  photons, respectively. \label{fig:escape_virial_void}}
\end{figure*}

\begin{figure*}[t]
\centering
\includegraphics[width=1.0\textwidth]{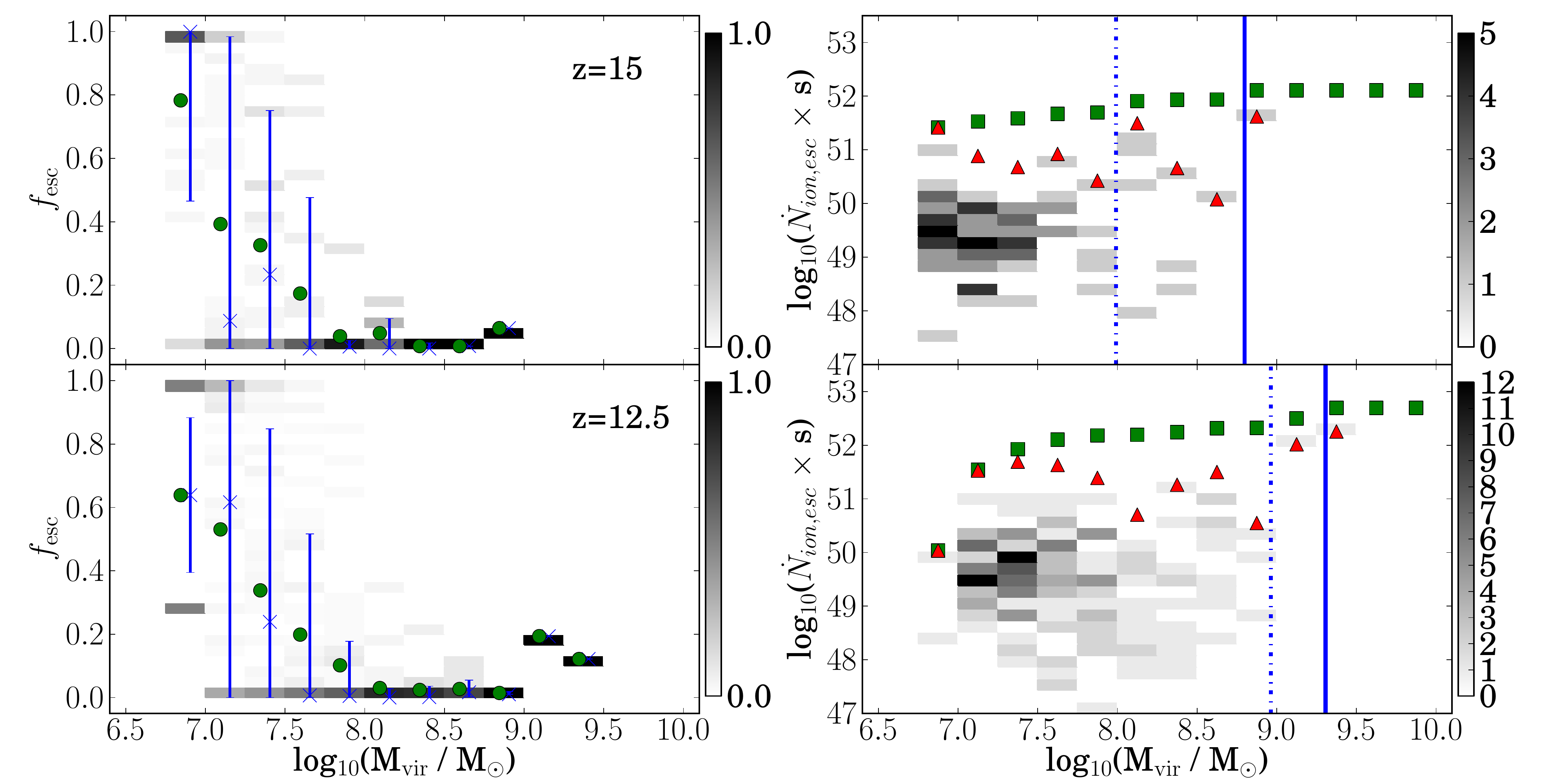}
\caption{Same as Figure \ref{fig:escape_virial_void}, but for the
  normal region at $z=15$ and 12.5.
\label{fig:escape_virial_normal}}
\end{figure*}

\begin{figure*}[t]
\centering
\includegraphics[width=1.0\textwidth]{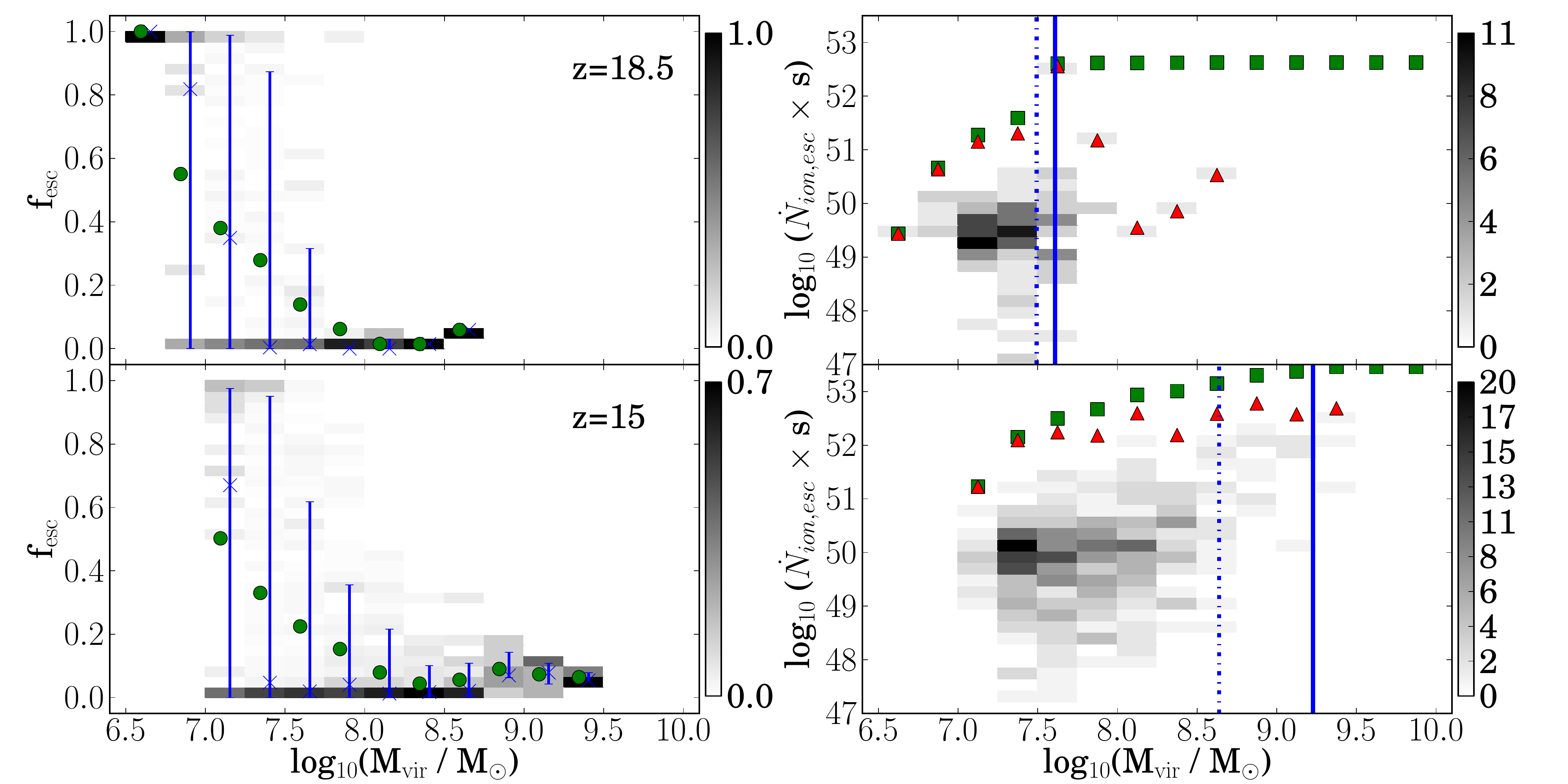}
\caption{Same as Figure \ref{fig:escape_virial_void}, but for the
  Rarepeak at $z=18.5$ and 15.
\label{fig:escape_virial_rarepeak}}
\end{figure*}

The UV escape fraction is notoriously difficult to calculate because
it depends on the current star formation rate and density and
ionization structure of the ISM, which can change rapidly from stellar
feedback.  Both UV radiation and SN explosions can create channels of
diffuse ionized gas, spanning from star formation regions to the IGM,
through which ionizing photons can escape.  In other sight lines,
however, there may be high density clumps or filaments that can absorb
a significant fraction of ionizing radiation.  Thus, the escape
fraction of each galaxy can vary in the temporal, spatial, and angular
domains.  Our simulations include radiation transport and SNe feedback
and can be used to calculate the UV escape fraction accurately in
post-processing.  Figure \ref{fig:halo_projects} shows the projections
of the baryon density and UV flux of the most massive halos in the
Void, Normal, and Rarepeak regions at $z = (8, 12.5, 15)$,
respectively.  These most massive galaxies in the Void ($M_{\rm vir} =
5.3 \times 10^9 \Ms;\; M_\star = 1.7 \times 10^8 \Ms$), Normal ($M_{\rm
  vir} = 2.7 \times 10^9 \Ms;\; M_\star = 5.7 \times 10^7 \Ms$), and
Rarepeak ($M_{\rm vir} = 1.9 \times 10^9 \Ms;\; M_\star = 4.8 \times
10^7 \Ms$) regions have escape fractions of about 0.042, 0.122 and
0.090, respectively.  It is apparent that the escaping radiation is
anisotropic, propagating through low-density channels in the porous
ISM \citep{Clarke02} and being absorbed by nearby dense gaseous
clumps.

\subsection{Method}

Because the Renaissance Simulations transports ionizing radiation that
is coupled to the energy and chemistry solver, the ionization
fractions stored in the datasets are accurate.  Thus, we can calculate
the escape fraction of all galaxies in a post-processing analysis.  We
only consider absorption by neutral hydrogen, which is the dominant
absorber in the energy range 10--50 eV, and neglect any other
absorbing species in the following analysis.  We use the same method
as \citet{Wise14} to calculate \fesc{}.  We briefly describe the
method next, leaving the details to the original paper.

The escape fraction of ionizing photons from a star particle to an IGM
component is dependent only on the optical depth $\tau = N_{\rm
  HI}(\vec{r}) \sigma(E)$, where $\vec{r}$ is the vector connecting
the two points.  Therefore, we can simply calculate the \hi{} column
density $N_{\rm HI}$ along various lines of sight and use the
photo-ionization cross-section $\sigma$ at $E = 21.6$ eV, which is the
average photon energy of the ionizing radiation considered in our
simulations.  We calculate 768 lines of sight from each star particle
to a sphere with radius $r_{\rm vir}$, centered on the halo center and
pixelated with HEALPix at level 3 \citep{HEALPix}.  We then compute
the associated optical depths and average $\exp(-\tau)$ over all
angles to calculate $f_{\rm i, esc}$ of a single star particle.  The
total UV escape fraction of a single galaxy is the luminosity-weighted
average of the escape fraction $f_{\rm i, esc}$ of all star particles
in the halo.

\subsection{Dependence on halo mass}

\begin{figure*}[t]
\centering
\includegraphics[width=1.0\textwidth]{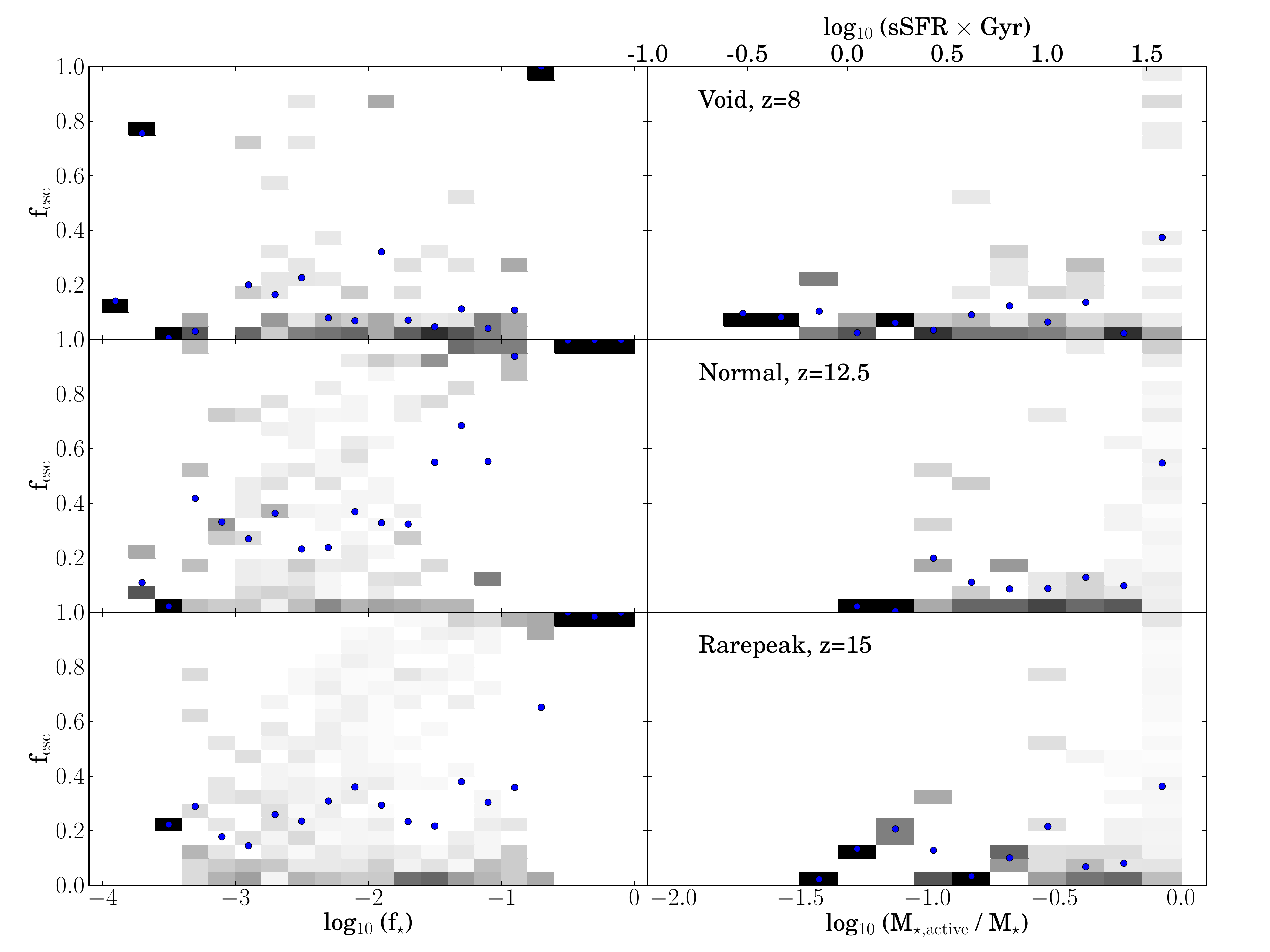}
\caption{The UV escape fraction as a function of stellar baryonic
  fraction f$_{\star}$ (left) and ratio of young ($<$ 20 Myr) UV
  radiative active stellar mass (right) of three simulations at $z=8$,
  12.5, and 15, respectively. The top axis of the right panel shows
  the corresponding specific star formation rate (sSFR) using the
  average SFR for the past 20 Myr. The blue circles show the average
  UV escape fraction. \label{fig:escape_fstar_ssfr}}
\end{figure*}

Figures \ref{fig:escape_virial_void}--\ref{fig:escape_virial_rarepeak}
show the distributions of \fesc{} and the total number of photons that
escape into the IGM for individual halos in the Void, Normal, and
Rarepeak regions at different redshifts, respectively.  These
distributions only include halos that have formed stars in the past 20
Myr -- in other words, the massive stars that produce nearly all of the ionizing
photons which have not yet ended their lives in supernovae.  First, we focus on the \fesc{}
distributions in the left panels of the Figures.  Their general trends
do not vary significantly with environment or redshift and follow
the same behavior as the results of \citet{Wise14}.  The low-mass
halos with $M_{\rm vir} \lsim 10^8~\Ms$ have a large scatter in
\fesc{} that is caused by the large variations in $f_\star$ and
$f_{\rm gas}$ from halo to halo.  Halos with low gas fractions and/or
high stellar baryonic fractions have the highest \fesc{} values.  The
smallest halos hosting metal-enriched star formation with $M_{\rm vir}
\simeq 10^7~\Ms$ have a median \fesc{} between 0.4 and 0.6, regardless
of redshift and region.  The median of the escape fraction steadily
decreases with increasing halo mass until it reaches $\sim$0.05 in the
range $M_{\rm vir} = 10^8 - 10^9~\Ms$.  However, it should be noted
that our most massive galaxies ($M_{\rm vir} > 10^9~\Ms$) in each
region have their \fesc{} values boosted to 0.1--0.2 as they
experience strong continuous star formation with SFR $\sim 1~\Ms
\unit{yr}^{-1}$ and sSFR $\sim 10^{-8} \unit{yr}^{-1}$. 

The right panels of Figures
\ref{fig:escape_virial_void}--\ref{fig:escape_virial_rarepeak} show
the distributions of the number of UV photons that escape into the IGM
from each halo as a function of halo mass.  We also show the total
number of escaping photons (red triangles) in each mass bin, along
with the cumulative number of escaping photons (green squares) below a
given halo mass in the panels.  To clearly denote which halos are
producing most of the escaping UV photons, we mark the halo mass in
which 50\% (dash-dotted lines) and 90\% (solid lines) of the ionizing
photons are produced in halos below these marked masses.  At early
times before any $10^9~\Ms$ halos form, the majority of the escaped
photons originate in halos below the atomic cooling threshold ($M_{\rm
  vir} \simeq 10^8~\Ms$).  However once more massive halos form at
later times, they contribute about half of the escaping UV photons to
the photon budget, even though our simulations only capture the
formation of a few halos with $M_{\rm vir} > 10^9~\Ms$ in each region.

\subsection{Dependence on recent star formation}

We have shown that the escape fraction depends on the halo mass, but
in principle it should also depend on the strength and timing of the
star formation because of the growth and breakout of \hii~regions from
young stellar populations.  Previous groups \citep{Wise09, Wise14,
  Kimm14} have found a correlation between the SFR and \fesc{} with
some delay as the ionization front propagates beyond the virial
radius.  Figure \ref{fig:escape_fstar_ssfr} plots the \fesc{}
distribution as a function of the stellar baryonic fraction $f_\star$
(left panels) and the stellar mass fraction in young stars (right
panels), or equivalently the sSFR, for all three regions at their
final redshifts.  To calculate the sSFR, we use the SFR averaged over
the past 20 Myr.  We find no general trends in \fesc{} with either
quantity and also observe a large scatter in \fesc{} in both.  However, there is an
indication of an upward trend at $f_\star \gsim 0.1$.  This scenario is rare and occurs
when the halo is partially photo-evaporated, leaving behind a dense
core and diffuse envelope.  Once stars form in this core, the
ionization front breaks out of the birth cloud and is largely
unimpeded by the photo-evaporated envelope.  This behavior also
manifests itself in the highest sSFR bin with several halos having
$f_{\rm esc} \gsim 0.5$.  With the exception of this extreme behavior
at large $f_\star$ values, the UV escape fractions are largely
independent of these star formation measures and vary substantially
between halos in our simulated sample of $\sim$2000 galaxies.

\subsection{Fractions of halos with star formation and photons escaped}

\begin{figure}[t]
\centering
\includegraphics[width=1.0\columnwidth]{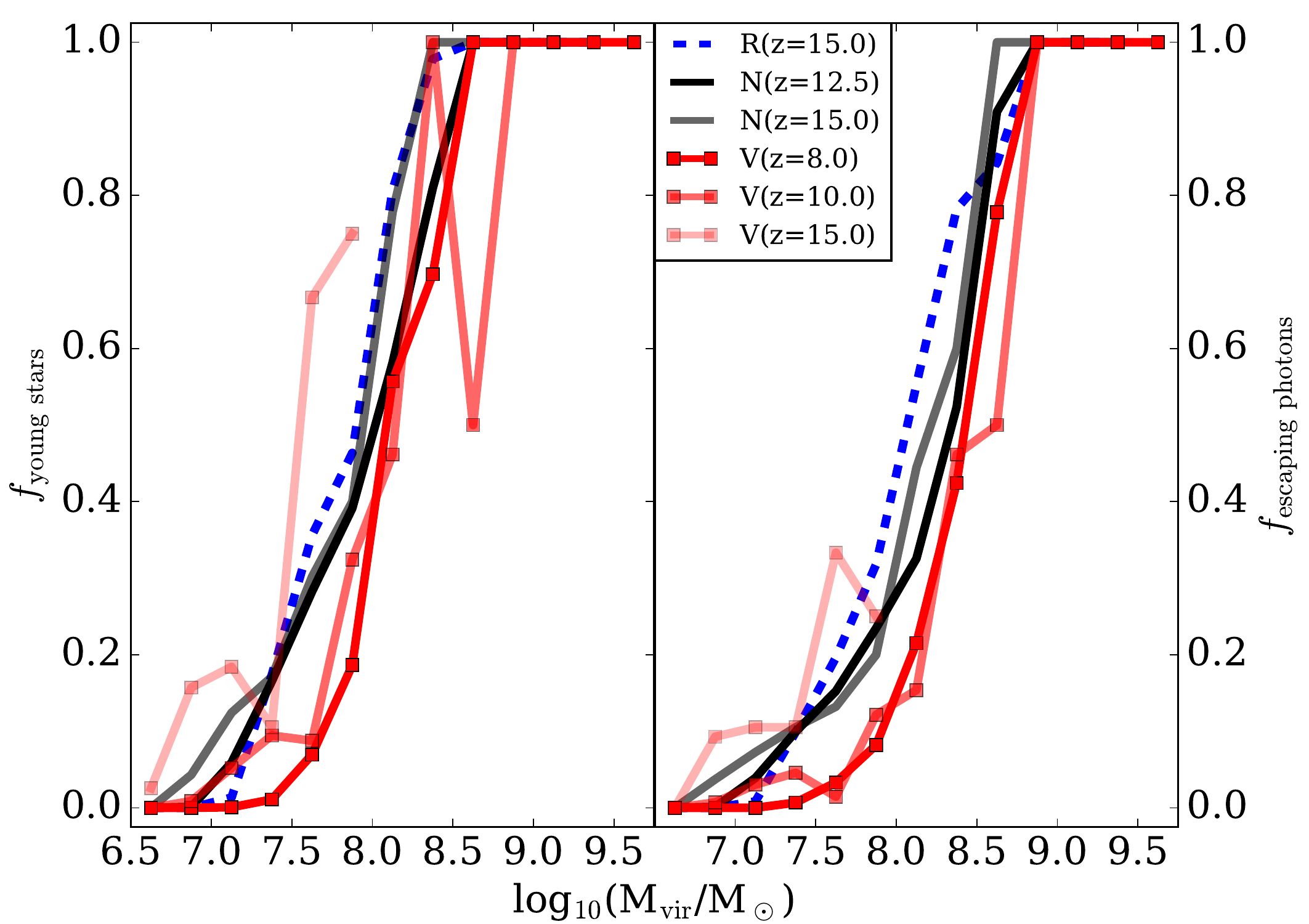}
\caption{Fractions of halos with active star formation (left; stars younger
  than 20 Myr) and photons escaped from their virial radius
  (right) as a function of the halo mass for the Rarepeak region (dashed
  blue) at $z=15$, Normal region (solid black) at $z = 15, 12.5$, and Void region
  (red solid with boxes) at $z = 15, 10, 8$.
  \label{fig:fraction_star_escape}}
\end{figure}

Gas in low mass halos are easily disrupted and expelled from the
shallow potential wells as the outflows generated by ionization fronts
and SNe easily exceed the escape velocity.  This feedback diminishes
the available fuel for star formation.  After some time elapses, the
cold gas reservoir is replenished by either cosmological infall or any
gas remaining in the halo cooling after the massive stars have died.
This cycle results in highly stochastic star formation in the
lower-mass halos in our simulations.  Figure
\ref{fig:fraction_star_escape} quantifies this behavior by plotting the
fraction of halos with active star formation (left panel), i.e., stars
younger than 20 Myr, and the fraction of halos with a non-zero \fesc{}
(right panel) as a function of halo mass.  We show these fractions for
all three regions at various redshifts.  Only 10\% of the $M_{\rm vir}
\sim 10^7~\Ms$ halos are hosting recent star formation at the times
shown, regardless of the redshift and region.  This fraction then
increases with halo mass to $\sim$50\% by $M_{\rm vir} \sim 10^8~\Ms$,
culminating in all halos with $M_{\rm vir} \ge 10^9~\Ms$ hosting
recent star formation.  These galaxy occupation fractions increase
similarly with halo mass in all regions at the times shown.  However
for a given region, this fraction decreases as time progresses at
$M_{\rm vir} \lsim 10^8~\Ms$ because of the increasing effects of
radiative feedback from more massive galaxies, effectively increasing
the filtering halo mass for efficient cooling and thus star formation
\citep[cf.][]{Gnedin00, Wise08_Reion}.  Lastly because ionizing
radiation originates from massive stars, the fraction of halos
generating escaping photons, shown in the right panel of Figure
\ref{fig:fraction_star_escape}, follow a similar trend as the stellar
occupation fraction.

\subsection{Comparison to other work}



Early efforts to constrain the escape fraction with simulations found
wildly varying \fesc{} values over a range of halo masses $10^7 -
10^{12}~\Ms$, ranging from less than 1\% to unity
\citep[e.g.][]{Fujita03, Razoumov06, Gnedin08_fesc, Wise09,
  Paardekooper11, Yajima11}.  However in the past few years with
higher resolution simulations that resolve star-forming clouds, there
has been some convergence, showing that low-mass ($\log M_{\rm
  vir}/\Ms \la 8$) halos have higher mean \fesc{} values $\ga 25\%$
\citep{Wise14, Paardekooper15} that then decrease with halo mass to
approach $\sim$5--15\% in more massive ($\log M_{\rm vir}/\Ms = 8 -
11$) galaxies at $z \ge 6$ \citep{Kimm14, Ma15}.  Using the star
formation surface density as a proxy to \fesc{}, \citet{Sharma16}
suggested that the brightest galaxies dominated the photon budget at
the end of reionization with 50\% of the photons originating from
galaxies with UV absolute magnitudes $M_{1500} \le (-18, -16.5)$ and
$\dot{M}_\star \ga (0.5, 0.1)~\Ms \unit{yr}^{-1}$ at $z = (6, 8)$,
respectively.

Our results are in agreement with these latest works with \fesc{}
decreasing with halo mass, given that $M_{\rm vir} \la 10^9~\Ms$.
Here we only compare the mean values, but in reality, it should be
noted that escape fractions greatly vary between galaxies and
temporally.  In the smallest galaxies, mean \fesc{} values are
$\sim$50\% in the range $10^7 - 10^{7.5}~\Ms$ with a large scatter
\citep[cf.][]{Wise14, Paardekooper15}, spanning the full range from
zero to unity, depending on previously discussed conditions.  This
then decreases to $\sim$0.05 at higher halo masses, agreeing with
\citet{Kimm14} and \citet{Ma15}.  The Void simulation at $z=8$
contains two galaxies with $M_\star \simeq 3 \times 10^{7}~\Ms$
contained in halos with $M_{\rm vir} = 2 \times 10^9~\Ms$ that have
escape fractions of 25\% that are undergoing strong starbursts.  These
galaxies may be analogs of the ones explored in the ``brightest
galaxies reionized the universe'' scenario posed by \citet{Sharma16}.
In the same data, we are also in agreement with \citeauthor{Sharma16},
where we both find that 50\% of the ionizing photons (see Figure
\ref{fig:escape_virial_void}) come from galaxies with a $\mathrm{SFR}
\ge 0.1~\Ms\unit{yr}^{-1}$ in halos $M_{\rm vir} \ge 10^9~\Ms$ at $z =
8$.



\section{Conclusions}
\label{sec:conclusions}
In this paper, we have quantified the properties and the UV escape
fraction of galaxies during the epoch of reionization, expanding upon
the work of \citet{Wise14} with a much larger sample of $\sim$2,000
galaxies and extending the relations to a maximum halo mass of $\sim
10^{9.5}~\Ms$.  To establish these relations, we have run and analyzed
the {\it Renaissance Simulations}, a suite of high-resolution
radiation hydrodynamics simulations of the first galaxies in three
different large-scale environments, each with a comoving volume of
$\sim$$200 \unit{Mpc}^3$, that follow Pop III star formation and the
transition to metal-enriched stars in thousands of halos.

In particular, we have analyzed these halos to determine any trends in
star formation and gas fraction with halo mass and environment.  Using
our large sample of first galaxies, we showed that their
characteristic properties and contribution to the photon budget of
reionization are mainly determined by their halo masses and are nearly
independent of the environment and redshift during the epoch of
reionization.  This finding validates the assumption of \citet{Wise14}
and \citet{Chen14} that the galaxy population during this epoch is
nearly invariant with redshift, which allowed them to increase their
effective galaxy sample size by combining datasets at different
redshifts into a single sample.

We found that galaxies with $M_{\rm vir} < 10^9~\Ms$ are consistent
with the high-resolution simulation results of \citet{Wise14}, which
only followed the formation of 32 galaxies by $z \sim 7$.  This result
is not unexpected because we use the same star formation and feedback
models; however, it does show that these models give a consistent
result even with a mass and spatial resolution that is a factor of 10 times
larger.  Given our larger volume, we are able to extend these
relations to halos with masses up to $10^{9.5}~\Ms$.  Equivalently,
our simulated UV galaxy LF now extends to $M_{1500} \simeq -18$ and
matches the abundances of the faintest galaxies observed at $z = 7-8$.
We find that our LF follow a similar power law seen in observations
\citep[e.g.][]{Bouwens15, Finkelstein15}, but extends to very faint
galaxies with $M_{1500} \simeq -12$ and flattens above this absolute
magnitude \citep{Wise14, OShea15}.

These larger galaxies provide some insight into the evolution of the
stellar sources of cosmic reionization.  We found that the decreasing
trend of \fesc{} with halo mass with a minimum of 0.05 between
$10^8-10^9~\Ms$; however the mean \fesc{} values stabilize around
0.1--0.2 at halo masses $M_{\rm vir} \gsim 10^9~\Ms$, which is
consistent with previous findings of \citet{Kimm14}.  Once halos reach
this mass, they are largely resistant to SN feedback and external
radiative feedback \citep[e.g.][]{Efstathiou92, ThoulWeinberg96,
  Okamoto08, Finlator11}.  They initially host relatively strong
(i.e., significant sSFR) and continuous star formation, and coupled with an escape
fraction in the range 0.1--0.2, they provide a large fraction
($\sim$50\%) of the photon budget of reionization at late times
\citep[see also][]{Sharma16}.  

We see that the faintest galaxies below the atomic cooling limit
($M_{\rm vir} \lsim 10^8~\Ms$) dominate the photon budget during the
initial stages of reionization, but then their star formation is
suppressed, leaving more massive galaxies to provide the majority of
UV radiation to complete reionization.  However, the escape fraction
alone does not tell the whole story.  Although low-mass galaxies
generate most of the ionizing photons at early times, the earliest
low-mass galaxies tend to form in overdense regions with high
recombination rates, resulting in smaller \hii~regions.  This effect
is apparent in the neutral fraction projections in Figure
\ref{fig:slice_rarepeak}, where the cosmological \hii~regions are
contained completely in the collapsing large-scale overdensity.
Although these early galaxies ionize a substantial mass fraction of
their local environment, they add little to the overall ionized
photons needed for complete reionization at $z \sim 6-7$.  We only see
significant growth of the volume- and mass-weighted ionization
fraction at late times -- in particular in the Void
region at $z = 8$.  Nevertheless, these faintest galaxies play an
important role in heating the surrounding IGM and shaping internal
gaseous galactic properties that affect subsequent galaxy formation.


\acknowledgements 
We thank an anonymous referee for a positive and insightful review
that improved the clarity of the paper.  This research was supported
by National Science Foundation (NSF) grant AST-1109243 to MLN. JHW
acknowledges support from NSF grants AST-1211626 and AST-1333360 and
Hubble theory grants HST-AR-13895 and HST-AR-14326.  KA was supported
by the NRF grant NRF-2014R1A1A2059811.  BWO was supported in part by
the sabbatical visitor program at the Michigan Institute for Research
in Astrophysics (MIRA) at the University of Michigan in Ann Arbor, by
NASA grants NNX12AC98G, NNX15AP39G, and by Hubble Theory Grants
HST-AR-13261.01-A and HST-AR-14315.001-A.  The simulation was
performed using \textsc{Enzo} on the Blue Waters operated by the
National Center for Supercomputing Applications (NCSA) with PRAC
allocation support by the NSF (award number ACI-0832662). Data
analysis was performed on the Gordon supercomputer operated for XSEDE
by the San Diego Supercomputer Center and on the Blue Waters
supercomputer. This research is part of the Blue Waters
sustained-petascale computing project, which is supported by the NSF
(award number ACI 1238993) and the state of Illinois. Blue Waters is a
joint effort of the University of Illinois at Urbana-Champaign and its
NCSA.  This research has made use of NASA's Astrophysics Data System
Bibliographic Services.  The majority of the analysis and plots were
done with \textsc{yt} \citep{yt_full_paper}.  \textsc{Enzo} and
\textsc{yt} are developed by a large number of independent researchers
from numerous institutions around the world. Their commitment to open
science has helped make this work possible.


\end{document}